\documentclass{article}
\usepackage{graphicx} 
\usepackage{authblk}

\usepackage{cite}
\usepackage{amsmath,amssymb,amsfonts}
\usepackage{algorithmic}
\usepackage{graphicx}
\usepackage{textcomp}
\usepackage{xcolor}
\def\BibTeX{{\rm B\kern-.05em{\sc i\kern-.025em b}\kern-.08em
    T\kern-.1667em\lower.7ex\hbox{E}\kern-.125emX}}

\usepackage{stmaryrd}
\usepackage{centernot}
\usepackage{enumerate}
\usepackage{url}
\usepackage{tikz}
\usepackage{pgf-umlsd}
\usepackage[linktoc=all,pdftex,pdfpagelabels,bookmarks,hyperindex,hyperfigures,hidelinks]{hyperref}

\input xy 
\xyoption{all}

\usepackage{comment}

\graphicspath{{figures/}}
\usepackage{caption}
\usepackage{subcaption}

\usepackage{relsize}
\usepackage{fancyvrb}
\fvset{numbersep=2pt, fontfamily=tt, fontsize=\smaller, frame=single}

\newcommand{\proto}{\caP}
\newcommand{\dlct}{\mathcal{D}}
\newcommand{\lingo}{\Lambda}

\usepackage{marginnote}
\usepackage[normalem]{ulem}
\newcommand{\editornote}[3]{%
    {\color{#3}%
        {\marginnote%
            [\color{#3}#2$\vartriangleright$]{\color{#3}$\vartriangleleft$#2}
        }%
        {#1}%
    }%
}
\newcommand{\victor}[1]{\editornote{#1}{Víctor}{blue}}
\newcommand{\jose}[1]{\editornote{#1}{Jose}{teal}}

\newcommand{\catherine}[1]{\editornote{#1}{Catherine}{red}}

\newtheorem{example}{Example}
\newtheorem{definition}{Definition}
\newtheorem{theorem}{Theorem}
\newtheorem{lemma}{Lemma}
\newtheorem{corollary}{Corollary}

\makeatletter
\newcommand*{\vneq}{%
  \mathrel{%
    \mathpalette\@vneq{=}%
  }%
}
\newcommand*{\@vneq}[2]{%
  \sbox0{\raisebox{\depth}{$#1\neq$}}%
  \sbox2{\raisebox{\depth}{$#1|\m@th$}}%
  \ifdim\ht2>\ht0 %
    \sbox2{\resizebox{\vneqxscale\width}{\vneqyscale\ht0}{\unhbox2}}%
  \fi
  \sbox2{$\m@th#1\vcenter{\copy2}$}%
  \ooalign{%
    \hfil\phantom{\copy2}\hfil\cr
    \hfil$#1#2\m@th$\hfil\cr
    \hfil\copy2\hfil\cr
  }%
}
\newcommand*{\vneqxscale}{1}
\newcommand*{\vneqyscale}{1}
\makeatother

\newcommand{\caP}{{\cal P}}

\title{Protocol Dialects as Formal Patterns: \\A Composable Theory of Lingos\\ Technical report}
\author[1]{Víctor García}
\author[1]{Santiago Escobar}
\author[2]{Catherine Meadows}
\author[3]{Jose Meseguer}
\affil[1]{Universitat Politècnica de València, Spain}
\affil[2]{Retired}
\affil[3]{University of Illinois at Urbana-Champaign, USA}
\date{\today}

\begin{document}

\maketitle

\begin{abstract}
Protocol \emph{dialects} are methods for modifying protocols that provide light-weight security, especially against easy attacks that can lead to more serious ones.  A \emph{lingo} 
is a dialect's key security component by making attackers unable to ``speak'' the lingo.  A lingo's ``talk'' changes all the time, becoming
a moving target for attackers.  We present several kinds of lingo transformations and compositions
to generate stronger lingos from simpler ones, thus making dialects more secure.

\end{abstract}

%
%

\section{Introduction}\label{sec:introduction}
Protocol dialects are methods for modifying protocols in order to provide a light-weight layer of security, especially against relatively easy attacks that could potentially be leveraged into more serious ones. The scenario is usually that of a network of mutually trusting principals, e.g. an enterprise network, that needs to defend itself from outside attackers. 
The most effective approach is to prevent outside parties from even  initiating a communication with group members, e.g. by requiring messages to be modified in some way unpredictable by the attacker.\footnote{By ``message'' we  mean a unit of information transferred by the protocol, and can mean anything from a packet to a message in the traditional sense. }

Protocol dialects are intended to be used as a first line of defense, not as a replacement for traditional authentication protocols.  In fact, they are primarily concerned with protecting against weak attackers that are trying to leverage off protocol vulnerabilities.  We can make use of this restricted scope, as well as the assumption that all parties in an enclave trust each other, to use an attacker model based on the on-path attacker model \cite{gogineniverify,ProtocolDialects-FormalPatterns}:  First, the attacker can read messages, but cannot destroy or replace them.  Second, principals  trust each other and share a common seed of a pseudo-random number generator that they can use to generate and share common  secret parameters.  
Because of the simplified threat model and trust assumption,
dialects can be designed to be simpler than full-scale authentication protocols.  Thus they can be expected to be prone to fewer implementation and configuration errors than full-scale protocols.  With this in mind, dialects can be used as a separate but very thin layer in the protocol stack, useful as a defense against attacks in case the main authentication protocol is misconfigured or implemented incorrectly.   Considering the dialect as a separate layer makes it easier to keep it simple, because it makes it possible to avoid dependencies on the layers it is communicating with.  In our work, we consider dialects that sit right above the transport layer in the TCP/IP model, while the protocol it is modifying sits just above the dialect in the application layer.  However, other locations are possible for the protocol being modified.  \ 


In \cite{ProtocolDialects-FormalPatterns} we introduced a system for specifying protocol dialects as \emph{formal patterns} that specify solutions to frequently occurring distributed system problems that are generic, executable, and come with strong formal guarantees.  Moreover, formal patterns can be further composed and instantiated to meet specific needs. The dialect pattern is divided into two components: the \emph{lingo}, which describes the actual message transformation, and the \emph{dialect}, which is responsible for managing lingos and interfacing with the underlying protocol.  Methods are also supplied for composing lingos, which allows for the creation of new  lingos and dialects from previous ones.  

We build on our work in \cite{ProtocolDialects-FormalPatterns} by applying it to the  design of  dialects. 
However, unlike  \cite{ProtocolDialects-FormalPatterns} we concentrate on the design of lingos instead of the design of dialects as a whole.  In order to facilitate this, instead of composing whole dialects as in \cite{ProtocolDialects-FormalPatterns},  we develop new formal patterns to create and compose new lingos from simpler ones, which we find to be a more flexible approach that allows better concentration on the problem at hand. These patterns are \emph{lingo transformations}, which map one or more lingos enjoying some properties to a new lingos enjoying some new properties.  In particular, we concentrate on lingos that are based on constantly changing secret parameters, which are generated by a  pseudorandom number generator shared throughout, itself parametric on a single shared secret, an approach to lingo design introduced by  Gogeneni et al. in \cite{gogineniverify}.   Our specifications are written in the Maude rewriting logic language, which allows for executable specifications and model checking rooted in logical theories, including both probabilistic and statistical techniques. Indeed, all examples shown in this paper are executable, and we provide executable Maude specifications in \url{https://github.com/v1ct0r-byte/Protocol-Dialects-as-Formal-Patterns}.

The paper is organized as follows.  Section~\ref{sec:preliminaries} introduces some preliminaries needed to read the paper. 
 In Section~\ref{sec:lingos} we begin by concentrating  on several types of lingo transformations that can be used to guarantee specific properties of lingos such as non-malleability, authentication, and \emph{f-checkability}, the checkability of correctness by the dialect  instead of the protocol.  In Section~\ref{sec:compositions} we deal with a more general type of lingo transformation: \emph{compositions}.  These include functional composition (composing lingos as functions) and horizontal composition  (choosing between different lingos probabilistically).  We define lingo composition and introduce several examples to demonstrate its usefulness, including showing  how  composition can be used to produce a lingo that is stronger than either of its components.   In Section~\ref{sec:dialects} we introduce a formal pattern for the dialects that use the lingos, completing the picture. Section~\ref{sec:conclusion} provides some open questions and concludes the paper.

\subsection{Related Work}\label{subsec:related-work}
Probably, the first work on dialects is that of Sjoholmsierchio et al. in \cite{sjoholmsierchio_software-defined_2019}.  In this work, the dialect is proposed as a variation of an open-source protocol to introduce new security measures or remove  unused code, and is applied to the OpenFlow protocol, introducing, among other things a defense against cipher-suite downgrade attacks on TLS 1.2.  This variation was independent of TLS, and was achieved by adding proxies, thus allowing for modification without touching the rest of the system and foreseeing the use of dialects as thin layers.  In later work by Lukaszewski and Xie \cite{lukaszewski2022towards} a layer 4.5 in the TCP/IP model was proposed for dialects.  In our own model we also make use of proxies for both sending and receiving parties, but stop short of proposing an additional official layer.

The threat model we use in this paper and in  \cite{ProtocolDialects-FormalPatterns} is similar to that used by Goginieni et al. in \cite{gogineniverify} and Ren et. al \cite{ren2023breaking}: an on-path attacker who is not able to corrupt any members of the enclave, as discussed earlier. 
We have also followed  \cite{ren2023breaking}'s suggestion to view dialects as composable protocol transformations.  In this paper, we simplify this approach by reducing dialect transformations to more basic lingo transformations, which are generic, compositional methods for producing new lingos (and thus new dialects) from old ones.

In  \cite{garcia-alfaro_mpd_2021,gogineni_framework_2022,gogineniverify} Mei, Gogineni, et al. introduce an approach to dialects in which the message transformation is updated each time using a shared pseudo-random number generator.  This means that the security of a transformation depends as much, and possibly more, on its unpredictability as it does on the inability of the attacker to reproduce a particular  instance of a transformation.  We have adopted and extended this approach.  In particular, we model a transformation as a parametrized function, or \emph{lingo}, that produces a new message using the original message and the parameter as input.  Both lingo and parameter can be chosen pseudo-randomly, as described in  \cite{ProtocolDialects-FormalPatterns}. 

None of the work cited above (except our own) applies formal design and evaluation techniques.   However, in \cite{talcott2024dialects} Talcott applies formal techniques to the study of dialects running over unreliable transport, such as UDP.   \cite{talcott2024dialects}  is complementary to our work in a number of ways.  First, we have been looking at dialects running over reliable transport, such as TCP.  Secondly, we concentrate on a particular attacker model, the on-path attacker with no ability to compromise keys or corrupt participants, and then explore ways of generating dialects and lingos that can be used to defend against this type of attacker.  In \cite{talcott2024dialects}, however, the dialects are relatively simple, but the attacker models are explored in more detail,  giving the attacker a set of specific actions that it can perform.   Like our model, the work in \cite{talcott2024dialects} is formalized in the Maude formal specification language, providing a clear potential synergy between the two models in the future.

 \subsection{Contributions of this paper}
\begin{enumerate}
\item We present a theory of lingos that  adds important new \emph{properties}, \emph{transformations} and \emph{composition operations} to
the notion of lingo introduced in \cite{ProtocolDialects-FormalPatterns}.  The main goal of this theory is to improve the security of protocol dialects.  
\item We show that these new lingo transformations and composition operations are \emph{formal patterns}  \cite{DBLP:conf/adt/Meseguer22} that guarantee
desired properties and replace and greatly simplify 
two earlier formal patterns for dialects in \cite{ProtocolDialects-FormalPatterns}.  
These lingo transformations provide new compositional techniques for creating new  lingos and dialects.  
\item We provide formal executable specifications in Maude \cite{maude-book} for lingos, their transformations and compositions, and dialects, available in a GitHub repository referenced in this paper. 
\end{enumerate}

\setcounter{section}{1}
\section{Preliminaries}\label{sec:preliminaries}
We summarized background on Maude and Actor systems relevant for this paper.

\noindent {\bf Maude} \cite{maude-book,duran_programming_2020} is a formal specification language and system based on rewriting 
logic~\cite{MeseguerTCS92,20-years}, a logic for specifying a wide range of concurrent systems, including, e.g.,
Petri nets \cite{stehr-meseguer-olveczky}, process calculi \cite{Verdejo-Marti-Oliet00}, object-based systems \cite{ooconc}, asynchronous hardware \cite{DBLP:journals/jlp/KatelmanKM12}, mobile ad hoc network protocols \cite{liu-etal-MANETS-JLAMP}, cloud-based storage systems \cite{cloud-chapter}, web browsers \cite{IE-maude-analysis}, concurrent programming languages \cite{meseguer-rosu-tcs}, distributed  control systems \cite{multirate-pals-SCP} and models of mammalian cell pathways \cite{pathways02,pathways04}.  A rewite theory is a triple $\mathcal{R}=(\Sigma,E,R)$,
where $(\Sigma,E)$ is an equational theory defining the system states as elements of the initial algebra
$T_{\Sigma/E}$, and $R$ is a collection of rewrite rules
of the form $t \rightarrow t'$ that describe the local concurrent transitions of the
concurrent system specified by $\mathcal{R}$.

\noindent {\bf Maude Functional (resp. System) Modules and Theories}. Rewriting Logic has Equational Logic as a sublogic. 
In Maude this corresponds to the sublanguage of \emph{functional modules and theories}.  
A  Maude \emph{functional module} is an equational theory $(\Sigma,E)$ delimited with keywords \texttt{fmod} and \texttt{endfm}
and having an \emph{initial algebra semantics}, namely, $T_{\Sigma/E}$.  It includes type and operator declarations
for $\Sigma$ (with the \texttt{sort}, resp. \texttt{op}, keywords), and equations $E$ declared with the \texttt{eq} keyword. Users can declare their own operator's syntax with '\texttt{\_}' denoting argument positions, e.g. \texttt{\_+\_} specifies a binary infix symbol. Maude binary operators can be specified with built-in equational attributes such as associativity (\texttt{assoc}), commutativity (\texttt{comm}) and identity (\texttt{id:}).
Likewise, a Maude \emph{functional theory} is an equational theory $(\Sigma,E)$ delimited by keywords \texttt{fth} and \texttt{endfth} 
with a \emph{loose semantics} (specifies all algebras satisfying the theory).
Similarly, a Maude \emph{system module} (resp. \emph{system theory}) is a rewrite theory $(\Sigma,E,R)$
delimited with keywords \texttt{mod} and \texttt{endm} (resp. \texttt{th} and \texttt{endth})
with an \emph{initial} (resp. \emph{loose}) model semantics.  The rules $R$ in a system module (resp, theory)
are introduced with the keyword \texttt{rl}.
See \cite{maude-book,maude-manual} for further details.

\noindent {\bf Maude Parameterized Modules}.
Maude supports \emph{parameterized functional modules}  having a \emph{free algebra semantics}.
They are parameterized by one of more functional theories.  For example, \verb+fmod MSET{X :: TRIV} endfm+
is a parameterized module with parameter theory \texttt{TRIV} having a single parameter type \texttt{Elt} and no
operations or equations that maps any chosen set $X$ of elements to the free algebra of (finite) multisets on the elements
of $X$.  The choice of an actual parameter $X$ is made by a Maude \emph{view} instantiating in this case 
\texttt{TRIV} to the chosen data type $X$.  For example, we can choose as elements the natural numbers
by a view \texttt{Nat} mapping \texttt{TRIV} to the functional module \texttt{NAT} of natural numbers
and mapping the parameter type (called a sort) \texttt{Elt} to the type (sort) \texttt{Nat} in \texttt{NAT}.
The semantics of the thus instantiated module \verb+fmod MSET{Nat}+ is the free algebra
of multisets with elements in the natural numbers.  
%
%
Likewise, \emph{parameterized system modules}  have a \emph{free model semantics}.
For example, a parameterized system module \verb+mod CHOICE{X :: TRIV} endm+ that imports 
the parameterized functional module \verb+fmod MSET{X :: TRIV} endfm+ can have a  rewrite rule
for non-deterministically choosing an element
in a multiset of elements.  Then, the instantiation \verb+mod CHOICE{Nat} endm+ provides non-deterministic
choice of an element in a multiset of natural numbers.

\noindent {\bf Actors} \cite{Agha86} model distributed systems in which  distributed objects communicate through asynchronous message passing. When an actor receives a message, it changes its state, may send new messages, and may create new actors. 
In \cite{DBLP:journals/pacmpl/LiuMOZB22,ProtocolDialects-FormalPatterns}, actors have been extended to 
\emph{generalized actor rewrite theories} (GARwThs). These are object-oriented rewrite theories 
specified in Maude as a special class of system modules
that satisfy natural requirements. 
The distributed states of a GARwTh (terms of sort \texttt{Configuration}) are multi-sets of objects (terms of sort \texttt{Object}) and messages (terms of sort \texttt{Msg}). Multiset union is modelled by an associative and commutative operator `\verb!_ _!' ( juxtaposition), with \texttt{null} is the empty multi-set. Communication protocols are typical GARwThs.

\subsection{Formal Semantics of MQTT} \label{subsec:MQTT}

MQTT ~\cite{HunkelerTS08} is a lightweight, publish-subscribe, protocol 
popular in devices with limited resources or network bandwidth. 
It requires a transport protocol such as TCP/IP that can provide ordered, lossless, bi-directional connections.  MQTT provides no security itself, and, although it can be run over secure transport protocols such as TLS, it is known to be commonly misconfigured \cite{hron20189mqtt}. This makes it an attractive application for protocol dialects
\cite{gogineniverify}. We use MQTT as the base protocol for the examples shown in the paper. 

\begin{example}\label{ex:MQTT-formal-semantics}
We formalize MQTT in Maude as a protocol example for Dialects. Specifically, we select to specify a subset of MQTT for illustration purposes. We give formal semantics to the following MQTT commands: connect, subscribe, unsubscribe, publish and disconnect. Furthermore, in our MQTT's symbolic model we have two classes of actors, \texttt{MqttClient} and \texttt{MqttBroker}. Class \texttt{MqttClient} defines three attributes: i) the broker it is connected to, ii) a list of commands to execute in order, and iii) a map between topic and value, to store the last value received on the subscribed topic. Class \texttt{MqttBrocker} has two attributes: i) a lists of peers and ii) a map between topics and the peers subscribed to it. With this two classes and set of commands we can build and represent environments with multiple clients and brokers, exchanging messages in a concurrent and asynchronous manner. 

As an example of the formal semantics of MQTT, we show the process of a client connecting to a broker with the actor rules of Figure~\ref{fig:mqtt-actor-rules}. First rule models the behaviour of an \texttt{MqttClient} with identifier \texttt{Me} sending a \texttt{connect} message to an \texttt{MqttBroker} with identifier \texttt{B}. Second rule represents what the \texttt{MqttBroker} will do upon receiving the \texttt{connect} message, i.e. add the \texttt{MqttClient} to it's set of peers and send back to the sender an acknowledgement message. With the third rule, the \texttt{MqttClient} processes any incoming acknowledgement message by setting its \texttt{peer} attribute with the senders identifier.

\begin{figure}[t]
\begin{Verbatim}
rl [mqtt/C/send-connect] :
    < Me : MqttClient | peer : nothing, cmdList : connect(B); CMDL, Atrs >
    =>
    < Me : MqttClient | peer : nothing, cmdList : CMDL, Atrs > 
    (to B from Me : connect(B)) .

rl [mqtt/B/accept-connect] : 
    < Me : MqttBroker | peers : Ps, Atrs > (to Me from O : connect(Me)) 
    =>
    < Me : MqttBroker | peers : insert(O, Ps), Atrs > (to O from Me : connack) .

rl [mqtt/C/recv-connack] : 
    < Me : MqttClient | peer : nothing, Atrs > (to Me from O : connack)
    =>
    < Me : MqttClient | peer : O, Atrs > .
\end{Verbatim}
\caption{Subset of actor rules from the MQTT Maude specification, formalizing the semantics of a client connecting to a broker.}
\label{fig:mqtt-actor-rules}
\end{figure}

The initial configuration, represented in Figure \ref{fig:mqtt-intial-conf}, has three objects: two clients (with identifiers \texttt{c1} and \texttt{c2} respectively) and one broker (with identifier \texttt{b}). Given the commands in the \texttt{cmdList} attribute of each client, what will happen is: 1) both clients will connect to the broker, 2) client \texttt{c1} will subscribe to topic \texttt{temperature} and, 3) client \texttt{c2} will publish value \texttt{34} on that same topic. In the end, client \texttt{c1} will receive a value published by client \texttt{c2} regarding the temperature, storing the last received value (34) in attribute \texttt{lastRecv}.
For the dialect examples in this work, we use this initial configuration (state). 

\begin{figure}[t]
\begin{Verbatim}
< c1 : MqttClient | peer : nothing, 
                    cmdList : (connect(b) ; subscribe("temp")), 
                    lastRecv : empty > 
< c2 : MqttClient | peer : nothing, 
                    cmdList : (connect(b) ; publish("temp", "34")), 
                    lastRecv : empty > 
< b : MqttBroker | peers : empty, subscribers : empty >
\end{Verbatim}
\caption{Initial MQTT configuration for experiments.}
\label{fig:mqtt-intial-conf}
\end{figure}
\end{example}

\section{Lingos}\label{sec:lingos}
A lingo is a data transformation $f$ between two data types
$D_{1}$ and $D_{2}$ with a one-sided inverse transformation $g$.  
The transformation $f$ is \emph{parametric} on a 
secret parameter value $a$ belonging to a parameter set $A$.  
For each parameter $a$, data from $D_{1}$ is transformed into data of $D_{2}$, which, using the same parameter $a$, can be transformed 
back into the original data from $D_{1}$ using the one-sided inverse $g$.
In all our applications
lingos will be used to transform the payload of a message 
in some protocol $\mathcal{P}$ with data type $D_1$ of payload
data.  $D_{2}$ will then be the data type of transformed payloads.
Such transformed payloads can then be sent, either as 
a single message or as a sequence of messages, to make
it hard for malicious attackers to interfere with 
the communication of honest participants, who are the
only ones knowing the current parameter $a \in A$.
The point of a lingo is that when such 
transformed messages are received by an honest
participant they can be transformed back using $g$ (if they were broken
into several packets they should first be reassembled) to get the original payload in $D_{1}$.

 Briefly, the set $D_1$ can be thought of as analogous to the plaintext space, the set $D_2$ as analogous to the ciphertext space in cryptographic systems, and the parameter set $A$ (also known as the secret parameter set) as analogous to the key space.
The elements of $A$ are generally required to be pseudorandomly generated from a key shared by enclave members so it can't be guessed in advance, and it is also generally updated with each use, so information gleaned from seeing one dialected message should not provide any help in breaking another.  However, it is not necessarily required that an attacker not be able to guess $a$ after seeing $f(d_1,a)$.  The only requirement is that  it is not feasible for the attacker  to compute $f(d_1,a)$ without having have been told that $a$ is the current input to $f(d_1,a)$.  This is for two reasons.  First the lingo is intended to provide authentication, not secrecy.  Secondly, the lingo is only intended to be secure against an on-path attacker \cite{ProtocolDialects-FormalPatterns} that can eavesdrop on traffic but not interfere with it. Thus, if an enclave member sends a message $f(d_1,a)$ the attacker, even if it learns $a$, can't remove $f(d_1,a)$ from the channel and replace it with $f(d'_1,a)$ where $d'_1$ is a message of the attacker's own choosing. It can send $f(d'_1,a)$ after $f(d_1,a)$ is sent, but $f(d'_1,a)$ will not be accepted if the secret parameter is changed each time the message is sent. The enforcement of the policy on the updating of parameters is the job of the dialect, which will be discussed in Section~\ref{sec:dialects}.


\begin{definition}[Lingo]\label{def:lingo}
A \emph{lingo} $\Lambda$ is a 5-tuple  $\Lambda = (D_1,D_2,A,f,g)$, where $D_1$, $D_2$ and $A$ are sets called, respectively, the \emph{input}, \emph{output}
and \emph{parameter} sets,
$f$, $g$ are functions $f : D_1 \times A \rightarrow D_2$,  $g : D_2 \times A \rightarrow D_1$ such that\footnote{The equality $g(f(d_1,a),a)=d_1$
can be generalized to 
an equivalence $g(f(d,a),a) \equiv d$. The generalization of lingos to allow for such message equivalence is left for future work.} 
$\forall d_1 \in D_1$, $\forall a \in A$, $g(f(d_1,a),a)=d_1$. 
We call a lingo \emph{non-trivial} iff $D_1$, $D_2$ and $A$
are non-empty sets. In what follows, all lingos considered will be
non-trivial.

Note that a lingo $\Lambda = (D_1,D_2,A,f,g)$ is just a three-sorted
$(\Sigma,E)$-algebra, with sorts\footnote{Note the slight abuse of notation:
in $\Sigma$, $D_{1}$, $D_2$ and $A$ are uninterpreted sort
\emph{names}, whereas in a given lingo
$\Lambda' = (D'_1,D'_2,A',f',g')$, such sorts are respectively interpreted
as \emph{sets} $D'_{1}$, $D'_2$ and $A'$ and, likewise, the uninterpreted
function \emph{symbols} $f$ and $g$ in $\Sigma$ are
interpreted as actual \emph{functions} $f'$ and $g'$ in a given lingo.}
$D_{1}$, $D_2$ and $A$, 
function symbols  $f : D_1 \times A \rightarrow D_2$ and $g : D_2 \times A \rightarrow D_1$, and $E$ the single $\Sigma$-equation $g(f(d_1,a),a)=d_1$.

This means that there is a natural notion of lingo homomorphism, namely,
a $\Sigma$-homomorphism.  Given lingos $\Lambda = (D_1,D_2,A,f,g)$
and $\Lambda' = (D'_1,D'_2,A',f',g')$, a \emph{lingo homomorphism}
$h : \Lambda \rightarrow \Lambda'$ is a triple of functions
$h=(h_1 , h_2 , h_3 )$ with $h_1 : D_1 \rightarrow D'_1$,
$h_2 : D_2 \rightarrow D'_2$ and $h_3 : A \rightarrow A'$, s.t.
$\forall d_1 \in D_1$, $\forall d_2 \in D_2$, $\forall a \in A$,
\[
h_2 (f(d_1 , a)) = f'(h_1 (d_1 ), h_3 (a)) \;\;\; \mathit{and} \;\;\;
h_1 (g(d_2 , a)) = g'(h_2 (d_2 ), h_3 (a)).
\]
\end{definition}


In Maude, an equational theory $(\Sigma,E)$ can be
specified as a \emph{functional theory}.  Therefore, the above notion of
lingo has the following natural specification in Maude:

\begin{Verbatim}
fth LINGO is
    sorts D1 D2 A . op f : D1 A -> D2 . op g : D2 A -> D1 .
    var d1 : D1 . var a : A . eq g(f(d1,a),a) = d1 .
endfth
\end{Verbatim}

Since lingos are a key component of dialects\cite{ProtocolDialects-FormalPatterns}, there is some extra information that a lingo must provide so a dialect can apply it. We easily extend the above theory to one with: i) a function called \texttt{param} for computing the corresponding value $a \in A$ from natural numbers, and ii) ingress and egress values denoting how many messages the lingo receives as input, and how many messages does it provide as output.

\begin{Verbatim}
fth PMLINGO is
    protecting NAT .
    including LINGO .
    op param : Nat -> A .
    ops ingressArity egressArity : -> NzNat .
endfth
\end{Verbatim}

Note the asymmetry between $f$ and $g$, in the lingo
definition, since we
do not have an equation of the form 
$\forall d_2 \in D_2$, $\forall a \in A$, $f(g(d_2,a),a)=d_2$. 
The reason for this asymmetry is that, given $a \in A$,
the set $\{f(d_{1},a) \mid d_1 \in D_1 \}$ may be
a \emph{proper} subset of $D_{2}$.  However, we show
in Corollary \ref{cor:corollary2} below that
the equation $f(g(d_2,a),a)=d_2$ does hold for any
$d_2 \in \{f(d_{1},a) \mid d_1 \in D_1 \}$.
Other results and proofs are included in Appendix~\ref{app:proofs}.
Of course, for the special case of
a lingo such that  $\forall a \in A$,
$D_2 = \{f(d_{1},a) \mid d_1 \in D_1 \}$,
the equation 
$\forall d_2 \in D_2$, $\forall a \in A$, $f(g(d_2,a),a)=d_2$
will indeed hold. 
In particular, the set equality
$D_2 = \{f(d_{1},a) \mid d_1 \in D_1 \}$
holds for all $a \in A$ in the following example.

\begin{example}[The XOR$\{n\}$ Lingo] \label{ex:xor-lingo} 
Let $\Lambda_{\mathit{xor}}\{n\}=(\{0,1\}^{n},\{0,1\}^{n},\{0,1\}^{n},\oplus,\oplus)$, with $\_\oplus\_ : \{0,1\}^{n} \times \{0,1\}^{n} \rightarrow \{0,1\}^{n}$ the bitwise exclusive or operation. 

Note that $\Lambda_{\mathit{xor}}\{n\}$ is \emph{parametric} on $n \in \mathbb{N}\setminus \{0\}$.  That is, for each choice of $n \geq 1$ we get a corresponding lingo. This is a common phenomenon: for many lingos, $D_1, D_2$ and $A$ are not fixed sets, but \emph{parametrised data types}, so that for each choice of their parameter we get a corresponding instance lingo.
\end{example}

We can specify in Maude the parametrised data type $\{0,1\}^{n}$
parametric on a non-zero bit-vector length $n$ and its associated \texttt{xor} function.  Since in Maude's
built-in module \texttt{NAT} of (arbitrary precision) natural
numbers there is a function \texttt{xor} that, given numbers
$n,m$, computes the number  $n \; \texttt{xor}\; m$ whose associated
bit-string is the bitwise
exclusive or of $n$ and $m$ when represented as bit-strings, we can define such a parametrised module as follows:
    
\begin{Verbatim}
fth NzNATn is protecting NAT . op n : -> NzNat [pconst] . endfth

fmod BIT-VEC{N :: NzNATn} is protecting NAT .
    sorts BitVec{N} BitStr{N} . subsort BitVec{N} < BitStr{N} .

    op [_] : Nat -> BitStr{N} [ctor] . vars N M : Nat .
    cmb [N] : BitVec{N} if N < (2 ^ N$n) .

    op _xor_ : BitStr{N} BitStr{N} -> BitStr{N} [assoc comm] .
    op _xor_ : BitVec{N} BitVec{N} -> BitVec{N} [assoc comm] .
    eq [N] xor [M] = [N xor M] .
    ...
endfm

view 8 from NzNATn to NAT is op n to term 8 . endv
\end{Verbatim}

We can easily define its instantiation for bytes and evaluate some expressions.
\begin{Verbatim}
fmod BYTE is
 protecting BIT-VEC{8} * (sort BitVec{8} to Byte) .
endfm
==========================================
reduce in BYTE : [3] xor [5] .
result Byte: [6]
==========================================
reduce in BYTE : [1000] xor [1] .
result BitStr{8}: [1001]
\end{Verbatim}

\noindent Note that,
exploiting the bijective correspondence between
the decimal notation and the bit-string notation,
the parametric subsort \texttt{BitVec\{N\}}
of \texttt{BitStr\{N\}}
is defined as the bitvectors of dimension $n$ by the
conditional membership axiom \verb@[i] : BitVec\{N\} if i < (2 ^ N$n)@.  For convenience,
when $n=8$ we have renamed the sort \texttt{BitVec\{8\}} to \texttt{Byte}.

\vspace{1ex}

Having defined \verb@BIT-VEC{N :: NzNATn}@,
we can now define the parametrised lingo
$\Lambda_{\mathit{xor}}=(\{0,1\}^{n},\{0,1\}^{n},\{0,1\}^{n},\oplus,\oplus)$
in Maude as follows:
    
\begin{Verbatim}
fmod IDLINGO{L :: LINGO} is
    op f' : L$D1 L$A -> L$D2 . op g' : L$D2 L$A -> L$D1 .
    var d1 : L$D1 .  var d2 : L$D2 .  var a : L$A .
    eq f'(d1,a) = f(d1,a) . eq g'(d2,a) = g(d2,a) .
endfm

view xorl{N :: NZNATn} from LINGO to BIT-VEC{N} is
    sort D1 to BitVec{N} . sort D2 to BitVec{N} . sort A to BitVec{N} .
    op f to _xor_ . op g to _xor_ .
endv

fmod XOR-L{N :: NZNATn} is protecting IDLINGO{xorl{N}} . endfm
\end{Verbatim}

As explained in Section~\ref{sec:introduction}, Lingos are transformations used by Dialects to transform the messages from, and to, the underlying object. The parametrized module \texttt{IDLINGO} is our approach, for illustration purposes, to show the execution of a lingo's \texttt{f} and \texttt{g} functions. For more information on Dialects and an example we refer the reader to Section~\ref{sec:dialects}.

The meaning of the parametrized module \verb@IDLINGO{L :: LINGO}@
is that of the \emph{identical lingo} obtained
by instantiating the theory \texttt{LINGO} with concrete data
types \texttt{D1}, \texttt{D2} and \texttt{A}, and concrete
functions \texttt{f} and \texttt{g}.  Recall that such instantiations
are achieved in Maude by means of a \emph{view}.  The only notational
difference is that we rename \texttt{f} and \texttt{g} by
\texttt{f'} and \texttt{g'}.  We then instantiate the theory
\texttt{LINGO} to   \verb@BIT-VEC{N}@ by means of a \emph{parametrized view}
\verb@xorl{N :: NZNATn}@.  Our desired parametrized lingo
$\Lambda_{\mathit{xor}}=(\{0,1\}^{n},\{0,1\}^{n},\{0,1\}^{n},\oplus,\oplus)$
is then the parametrized module \verb@XOR-LINGO{N :: NZNATn}@. We can then instantiate it to bytes as follows:

\begin{Verbatim}
fmod BYTE-LINGO is
    protecting XOR-L{8} * (sort BitVec{8} to Byte) .
endfm
==========================================
reduce in BYTE-L : f'([3], [5]) .
result Byte: [6]
==========================================
reduce in BYTE-L : g'([3], [5]) .
result Byte: [6]
\end{Verbatim}

\begin{example}[XOR-BSEQ Lingo for Bit-sequences] \label{ex:XOR-BSeq-Lingo}
Yet another variation the same theme is to
define the $\Lambda_{xor.\mathit{BSeq}}$
lingo, whose elements are
\emph{bit sequences} of arbitrary length.  
This can be easily done
in Maude as follows:

\begin{Verbatim}
view xor-seq-l from LINGO to NAT is
    sort D1 to Nat . sort D2 to Nat . sort A to Nat .
    op f to _xor_ . op g to _xor_ .
endv
\end{Verbatim}
\end{example}

And instantiate \texttt{IDLINGO} to evaluate some expressions.
\begin{Verbatim}
fmod XOR-BSEQ-LINGO is
    protecting IDLINGO{xorl-Nat} .
endfm
==========================================
reduce in XOR-BSEQ-LINGO : f'(3, 5) .
result NzNat: 6
==========================================
reduce in XOR-BSEQ-LINGO : g'(6, 5) .
result NzNat: 3
\end{Verbatim}

\vspace{2ex}

Note that \texttt{BIT-VEC\{N\ ::\ NZNATn\}} 
is not the only possible parametrised data type on which a
lingo based on the $\mathit{xor}$ operation could be based.  A different
parametrised data type can have as elements the \emph{finite subsets} of
a parametrised data type of finite sets, whose elements
belong to a parameter set $D$, and where the $\mathit{xor}$
operation is interpreted as the \emph{symmetric difference}
of two finite subsets of $D$.  A finite subset, say, $\{a,b,c\} \subseteq D$,
with $a,b,d \in D$ different elements, can be represented
as the expression $a \; \mathit{xor} \; b \; \mathit{xor} \; c$.
Here is the parametrised module
defining such a parametric data type and its instantiation
for $D$ the set of quoted identifiers (a
built-in module \texttt{QID} in Maude with sort \texttt{Qid}).

\begin{Verbatim}
fmod XOR-SET{D :: TRIV} is
    sort Set . subsort D$Elt < Set .
    op zero : -> Set . op _xor_ : Set Set -> Set [ctor assoc comm] .
    var S : Set . eq S xor S = zero . eq zero xor S = S .
endfm

view Qid from TRIV to QID is sort Elt to Qid . endv
\end{Verbatim}

We can use both the  functional module and the  view to evaluate some expressions.
\begin{Verbatim}
fmod XOR-QID is
    protecting XOR-SET{Qid} .
endfm
==========================================
reduce in XOR-QID : ('a xor 'b xor 'c xor 'd) xor 
('c xor 'd xor 'e xor 'f) .
result Set: 'a xor 'b xor 'e xor 'f
==========================================
reduce in XOR-QID : 'b xor 'c xor 'b xor 'b xor 'd .
result Set: 'b xor 'c xor 'd
\end{Verbatim}

\noindent Note the isomorphism between vector and the finite set
representations:  If $|D|=n$, then the power set $\wp(D)$ with
symmetric difference is isomorphic to the function set
$[D \rightarrow \{0,1\}]$ with pointwise $\mathit{xor}$
of predicates $p,q \in [D \rightarrow \{0,1\}]$, i.e., 
$p \; \mathit{xor} \; q = \lambda d \in D.\; p(d) \;\mathit{xor} \; q(d)$,
which is itself isomorphic to $\{0,1\}^{n}$ with pointwise $\mathit{xor}$.

\vspace{2ex}

To obtain the corresponding parametrised $\mathit{xor}$ lingo
in its set representation we proceed in a way entirely analogous
to our Maude specification of $\Lambda_{\mathit{xor}}\{n\}$:

\begin{Verbatim}
view xorlingo{D :: TRIV} from LINGO to XOR-SET{D} is
    sort D1 to Set . sort D2 to Set . sort A to Set .
    op f to _xor_ . op g to _xor_ .
endv
\end{Verbatim}

And instantiate \texttt{IDLINGO} to evaluate some expressions.
\begin{Verbatim}
fmod LINGO-XOR{D :: TRIV} is
    protecting IDLINGO{xorlingo{D}} .
endfm
fmod QID-LINGO is
    protecting LINGO-XOR{Qid} .
endfm
==========================================
reduce in QID-LINGO : 
f'('a xor 'b xor 'c xor 'd, 'c xor 'd xor 'e xor 'f) .
result Set: 'a xor 'b xor 'e xor 'f
==========================================
reduce in QID-LINGO : 
g'('a xor 'b xor 'c xor 'd, 'c xor 'd xor 'e xor 'f) .
result Set: 'a xor 'b xor 'e xor 'f
\end{Verbatim}

\begin{example}[Divide and Check (D\&C) Lingo] \label{ex:d&c-lingo-old}\label{ex:d&c-lingo}
$\Lambda_{D\&C} = (\mathbb{N},\mathbb{N} \times \mathbb{N},\mathbb{N},
f, g)$, given $\forall n, a, x, y \in \mathbb{N}$ then:
\begin{itemize}
 \item $f(n,a) = (quot(n +(a+2),a +2),rem(n + (a+2),a+2))$
\item $g((x,y),a) = (x \cdot (a+2)) + y -(a+2)$
\end{itemize}
where $\mathit{quot}$ and $\mathit{rem}$ denote the quotient
and remainder functions on naturals.
\end{example}

\noindent The idea of the $\Lambda_{D\&C}$ lingo is quite simple.  Given a parameter 
$a \in \mathbb{N}$,
an input number $n$ is transformed into a
pair of numbers $(x,y)$: the quotient $x$ of $n+a+2$ by $a+2$ (this makes sure that
$a+2 \geq 2$), and the remainder $y$ of $n+a+2$ by $a+2$.  The meaning of
``divide'' is obvious.  The meaning of ``check''
reflects the fact that a receiver of a pair $(x,y)$ who knows $a$
can check $x = quot((x \cdot (a+2))+y,a+2))$
and $y = rem((x \cdot (a+2))+y,a+2))$, giving some assurance
that the pair $(x,y)$ was obtained from
$n = (x \cdot (a+2)) + y -(a+2)$ and
has not been tampered with.  This
is an example of an $f$-\emph{checkable} lingo (see Section~\ref{f-check-subsection}), whereas no such check is possible for any of the
isomorphic versions of the $\Lambda_{\mathit{xor}}$ lingo
(see again Section~\ref{f-check-subsection}). 
The lingo $\Lambda_{D\&C}$ can be formally specified in Maude as follows:

\begin{Verbatim}
fmod NAT-PAIR is protecting NAT .
    sort NatPair .
    op [_,_] : Nat Nat -> NatPair [ctor] . ops p1 p2 : NatPair -> Nat .
    vars n m : Nat . eq p1([n,m]) = n . eq p2([n,m]) = m .
endfm

view D&C-ling from LINGO to NAT-PAIR is
    sort D1 to Nat . sort D2 to NatPair . sort A to Nat .
    var n : D1 . var P : D2 . var a : A .
    op f(n,a) to term [(n+(a+2)) quo (a+2), (n+(a+2)) rem (a+2)] .
    op g(P,a) to term sd(((p1(P) * (a+2)) + p2(P)), (a+2))  .
endv
\end{Verbatim}

We instantiate \texttt{IDLINGO} to evaluate some expressions.
\begin{Verbatim}
fmod D&C-LINGO is
    protecting IDLINGO{D&C-ling} .
endfm
==========================================
reduce in D&C-LINGO : f'(13, 3) .
result NatPair: [3, 3]
==========================================
reduce in D&C-LINGO : g'([3, 3], 3) .
result NzNat: 13
==========================================
reduce in D&C-LINGO : g'(f'(13, 3), 3) .
result NzNat: 13
\end{Verbatim}

\subsection{Basic Properties of Lingos}

\noindent This section proves several useful properties of lingos.

\begin{lemma}\label{lem:lemma1}
Let $\Lambda=(D_1,D_2,A,f,g)$ be a lingo. Then, $\forall d_1, d'_1 \in D_1$, $\forall a \in A$ $f(d_1, a) = f(d'_1, a) \Rightarrow d_1 = d'_1$. 
\end{lemma}

\noindent {\bf Proof}:
Applying $g$ on both sides of the condition, by the
definition of lingo we get,
$d_1 = g(f(d_1,a),a) = g(f(d'_1,a),a) = d'_1$. $\Box$

\begin{lemma}\label{lem:lemma2}
Let $\Lambda=(D_1,D_2,A,f,g)$ be a lingo. Then, $\forall d_1 \in D_1$, $\forall d_2 \in D_2$, $\forall a \in A$, $d_2 = f(d_1,a) \Rightarrow d_1 = g(d_2,a)$.
\end{lemma}

\noindent {\bf Proof}:  Applying $g$ to both sides of the
condition, by the
definition of lingo we get,
$g(d_2,a) = g(f(d_1,a),a) = d_1$.  $\Box$.

\begin{corollary}\label{cor:corollary1}
Let $\Lambda=(D_1,D_2,A,f,g)$ be a lingo. Then, $\forall d_1 \in D_1$, $\forall d_2 \in D_2$, $\forall a \in A$, $d_2 = f(d_1, a) \Rightarrow d_2 = f(g(d_2, a),a)$. 
\end{corollary}

\noindent {\bf Proof}:
By Lemma \ref{lem:lemma2}, $d_1 = g(d_2,a)$. Thus, $d_2 = f(d_1,a) = f(g(d_2,a),a)$. $\Box$

\begin{corollary}\label{cor:corollary2}
Let $\Lambda=(D_1,D_2,A,f,g)$ be a lingo. Then, $\forall d_2 \in D_2$,  $\forall a \in A$, $((\exists d_1 
\in D_1, \; f(d_1, a) = d_2) \Rightarrow d_2 = f(g(d_2, a),a))$. 
\end{corollary}

\noindent {\bf Proof}: By Lemma \ref{lem:lemma2} we have,
$\forall d_2 \in D_2,\forall a \in A, (\forall d_1 \in D_1, d_2 = f(d_1,a) \Rightarrow d_2 = f(g(d_2,a),a)).$
which means
$\forall d_2 \in D_2,\forall a \in A, (\forall d_1 \in D_1, \;
\neg(d_2 = f(d_1,a)) \vee d_2 = f(g(d_2,a),a))$
which by $d_1$ not a free variable in $d_2 = f(g(d_2,a),a)$
is equivalent to
$\forall d_2 \in D_2,\forall a \in A, ((\forall d_1 \in D_1, \;
\neg(d_2 = f(d_1,a))) \vee d_2 = f(g(d_2,a),a))$
which by duality of quantifiers is equivalent to
$\forall d_2 \in D_2,\forall a \in A, (\neg (\exists d_1 \in D_1, \;
d_2 = f(d_1,a)) \vee d_2 = f(g(d_2,a),a))$
which means
$\forall d_2 \in D_2,\forall a \in A, ((\exists d_1 \in D_1, \;
d_2 = f(d_1,a)) \rightarrow d_2 = f(g(d_2,a),a))$
as desired. $\Box$

\vspace{1ex}

Corollary \ref{cor:corollary3} below shows that
users of a lingo can check if a received payload $d_2 \in D_2$ 
is of the form $d_2 = f(d_1 , a)$ for 
some $d_1 \in D_1$ and
$a \in A$
the current parameter, i.e., if it was generated
from some actual $d_1 \in D_1$ as $d_2 = f(d_1,a)$.
We call a $d_2 \in D_2$ such that $d_2 = f(d_1,a)$ for some
$d_1 \in D_1$ \emph{compliant} with parameter $a \in A$.
As shown in Corollary \ref{cor:corollary3}
compliance with $a$ can be checked by checking the equality $f(g(d_2,a),a) = d_2$ for any given $d_2 \in D_2$ and $a \in A$.

\begin{corollary}\label{cor:corollary3}
Let $\Lambda=(D_1,D_2,A,f,g)$ be a lingo. Then, $\forall d_2 \in D_2$,  $\forall a \in A$, $\exists d_1, (f(d_1, a) = d_2) \Leftrightarrow d_2 = f(g(d_2, a),a)$. 
\end{corollary}

\noindent {\bf Proof}:
The $(\Rightarrow)$ implication follows
from Corollary \ref{cor:corollary2}.
The $(\Leftarrow)$ implication follows
by choosing 
$d_{1} =g(d_2, a)$.  $\Box$

\subsection{f-Checkable Lingos} \label{f-check-subsection}


In this section, we discuss lingos where validity of its generated payloads can be checked by the dialect. We develop a lingo transformation that 
substantially reduces the probability an attacker successfully forging a message.



\begin{definition}[f-Checkable Lingo]  A Lingo $\Lambda=(D_1,D_2,A,f,g)$ is called
$f$-\emph{checkable} iff $\forall a \in A$ $\exists d_2 \in D_2$
s.t. $\not\exists \; d_1 \in D_1$ s.t. $d_2 = f(d_1,a)$. The 'f' in $f$-Checkable Lingo is an uninterpreted function symbol, but once an interpretation is given (i.e., through a view), f becomes interpreted as the function specified by the interpretation. 
\end{definition}

As just explained above, $\Lambda_{\mathit{xor}}$ is not $f$-checkable.
More generally, call a lingo $\Lambda=(D_1,D_2,A,f,g)$
\emph{symmetric} iff $(D_2,D_1,A,g,f)$ is also a lingo.
Obviously, $\Lambda_{\mathit{xor}}$ is symmetric.  Any
symmetric lingo is \emph{not} $f$-checkable, since 
any $d_2 \in D_2$
satisfies the equation $f(g(d_2,a),a) = d_{2}$.

\begin{example} \label{ex:f-Ckeckable-lingo} $\Lambda_{D \& C}$ is $f$-checkable.  Indeed,
for each $a \in \mathbb{N}$, any $(x,y)$ with $y \geq a +2$
cannot be of the form $(x,y) = f(n,a)$ for any $n \in \mathbb{N}$.
This is a cheap, easy check.  The actual check that a $d_1$ exists,
namely, the check $(x,y) = f(g((x,y),a),a)$ provided by Corollary \ref{cor:corollary3},
was already explained when making explicit the meaning of "C" in 
$\Lambda_{D \& C}$.
\end{example}




\begin{theorem}\label{theorem:trasnform-fCheckable}
    Let $\Lambda=(D_1,D_2,A,f,g)$ be a lingo, where
    we assume that $D_1$, $D_2$ and $A$ are computable data-types and $|D_{1}| \geq 2$
    %
    %
    Then, $\Lambda^{\sharp}=(D_1,{D_2}^2,A \otimes A,f^{\sharp},g^{\sharp})$ is an
    $f$-checkable lingo, where:
    \begin{itemize}
        \item $A \otimes A = A \times A \setminus id_A$, with $id_A = \{(a,a) \in A^2 | a \in A\}$
        \item $f^{\sharp}(d_1, (a,a')) = (f(d_1,a), f(d_1,a'))$
        \item $g^{\sharp}((d_2,d'_2), (a,a')) = g(d_2,a)$
    \end{itemize}
\end{theorem}

%
\noindent {\bf Proof}:
We first prove that 
$\Lambda^{\sharp}=(D_1,{D_2}^2,A \otimes A,f^{\sharp},g^{\sharp},\mathit{comp}^{\sharp})$ satisfies the lingo equation
$g(f(d_1,a),a)= d_1$.  Indeed,
\[
g^{\sharp}(f^{\sharp}(d_1,(a,a')),(a,a')) = g^{\sharp}((f(d_1,a), f(d_1,a')),(a,a')) = g((f(d_1,a), a) = d_1
\]
as desired.

\vspace{1ex}

\noindent To prove that, in addition,
$\Lambda^{\sharp}=(D_1,{D_2}^2,A \otimes A,f^{\sharp},g^{\sharp})$
is an $f$-checkable lingo, note that, since $|D_{1}| \geq 2$,
for each $(a,a') \in A \otimes A$ we can choose $d^{\sharp}_{2} = 
(f(d_1 ,a),f(d'_1,a'))$ with $d_1 \not= d'_1$.
Then, there is no $d''_1 \in D_1$
such that $d^{\sharp}_{2} = f^{\sharp}(d''_1,(a,a'))$,
since this would force $(f(d''_1 ,a),f(d''_1,a'))=(f(d_1 ,a),f(d'_1,a'))$,
which by Lemma \ref{lem:lemma1} would force $d''_1 = d_1 = d'_1$, which
is impossible.
$\Box$



\noindent The lingo transformation $\Lambda \mapsto \Lambda^{\sharp}$
has a simple specification in Maude 
as the following parametrised module (note that $A \otimes A$ is here denoted by \verb+APair#+):

\begin{Verbatim}
fmod LINGO#{L :: PMLINGO} is protecting INITIAL-EQUALITY-PREDICATE .
    sorts APair# APair D2Pair . subsort APair# < APair .
    
    op a[_,_] : L$A L$A -> APair [ctor] . 
    op d2[_,_] : L$D2 L$D2 -> D2Pair [ctor] .
    
    vars a a' : L$A .  var d1 : L$D1 . vars d2 d2' : L$D2 .
    cmb a[a,a'] : APair# if a .=. a' = false .

    op f# : L$D1 APair# -> D2Pair .
    eq f#(d1,a[a,a']) = < f(d1,a),f(d1,a') > . 
    op g# : D2Pair APair# -> L$D1 .
    eq g#(d2[d2,d2'],a[a,a']) = g(d2,a) .
    ...
endfm        
\end{Verbatim}

\begin{example}[$f$-checkable transformed version of 
$\Lambda_{\mathit{xor}}\{n\}$] \label{ex:edxor}
The lingo transformation $\Lambda \mapsto \Lambda^{\sharp}$
maps $\Lambda_{\mathit{xor}}\{n\}$ in Example \ref{ex:xor-lingo}
to the $f$-checkable lingo
$\Lambda_{\mathit{xor}}^{\sharp}\{n\}=(\{0,1\}^{n},\{0,1\}^{n}
\times \{0,1\}^{n},\{0,1\}^{n} \otimes \{0,1\}^{n},\oplus^{\sharp},
\oplus^{\sharp})$.  
\end{example}

\vspace{2ex}

\noindent The lingo $\Lambda_{\mathit{xor}}^{\sharp}\{n\}$
and it instance for bytes $\Lambda_{\mathit{xor}}^{\sharp}\{8\}$
are both obtained by instantiating the parametrised module
\verb@LINGO#{L :: PMLINGO}@ as follows:

\begin{Verbatim}
fmod XOR-LINGO#{N :: NzNATn} is protecting LINGO#{xorpml{N}} . endfm

fmod BYTE-LINGO# is 
    protecting XOR-LINGO#{8} * (sort BitVec{8} to Byte) . 
endfm
\end{Verbatim}

And evaluate some expressions.
\begin{Verbatim}
reduce in BYTE-LINGO# : f#([3], [[5], [7]]) .
result D2Pair: d2[[6], [4]]
==========================================
reduce in BYTE-LINGO# : 
g#(f#([3], [[5], [7]]), [[5], [7]]) .
result Byte: [3]
\end{Verbatim}

Suppose the probability of an attacker's successfully forging a message in the original lingo is $P_F$.  The probability of the attacker passing the $f$-checkability test in the transformed lingo is $1/2^n$.   Thus the probability of the attacker's successfully forging a message is the transformed lingo is $P_F/2^n$.

\subsection{Malleable Lingos}\label{subsubsec:comp-malleable-lingos}

\noindent A cryptographic function is called \emph{malleable} 
if an intruder, \emph{without knowing the secret key},
can use an existing encrypted message to generate another
message also encrypted with the same secret key.
In a similar way, if a lingo $\Lambda$ is what below we call
\emph{malleable}, an intruder can disrupt the communication
between an honest sender \emph{Alice} and an honest receiver 
\emph{Bob}
in a protocol $\mathcal{P}$ whose messages are modified
by means of a
lingo $\Lambda$\footnote{I.e., honest participants that use a 
\emph{dialect} $\mathcal{D}_{\Lambda}(\mathcal{P})$
based on $\mathcal{P}$ and
$\Lambda$ to modify their messages
(see \S \ref{sec:dialects}).} by producing
a message \emph{compliant} with
a secret parameter $a \in A$ of
$\Lambda$ 
supposedly sent from \emph{Alice} to 
\emph{Bob} (who share the secret parameter $a$),
 but actually sent by the intruder.
 
To see how malleability can become an issue for lingos, consider a variant of the xor lingo in which the same random bit-string is applied to two messages instead of one.  Thus the attacker will see $d_1 \oplus a$, then $d'_1 \oplus a$. Although the attacker may be able to find $a$ after seeing both messages, it will not be able to replace either one with a $d''_1 \oplus a$ of its own choosing, since it can't interfere with messages in flight.  But suppose that it has a good idea that the first byte of the first message will be $y$.  After it has seen $d_1 \oplus a$, it can then send $y;0^n \oplus z;0^n \oplus d_1 \oplus a$, thus sending a message identical to $d_1$ except that its first byte is $z$.  Note that the attacker can do this \emph{without} knowing $a$.  Thus it is not the fact that $a$ may derivable from the two messages that causes a security vulnerability, but the fact that exclusive-or is malleable. Here is the
definition.

\begin{definition}[Malleable Lingo] Let $\Lambda=(D_1,D_2,A,f,g)$ 
be a lingo which has been formally specified as 
a functional module in Maude by an equational theory
$(\Sigma_{\Lambda},E_{\Lambda})$ enjoying the required
executability conditions, so that its data types and
functions are all computable.  $\Lambda$ is called
\emph{malleable} iff there exists a $\Sigma_{\Lambda}$-term
$t(x,y)$ of sort $D_2$
with free variables $x,y$ of respective sorts $D_2$ and $A$
 and a computable function $r:A \rightarrow A$ such that
 $\forall d_1 \in D_1, \; \forall a,a' \in A,$
\begin{enumerate}
    \item $f(d_1,a) \not= t(f(d_1,a),r(a'))$, where, by convention,
  $t(f(d_1,a),r(a'))$  abbreviates the substitution instance
  $t\{x \mapsto f(d_1,a),y \mapsto r(a')\}$

  \item $\exists d'_1 \in D_1$ such that $t(f(d_1,a),r(a')) = f(d'_1,a)$.
\end{enumerate}
Note that, by (2) and the fact that $\Lambda$ is a lingo,
$d'_1 = g(t(f(d_1,a),r(a')),a)$.
\end{definition}

\noindent That is, if $a \in A$ is the current secret parameter 
that an honest sender \emph{Alice} is using to send $d_1$ 
(modified as $f(d_1,a)$)
to honest
receiver \emph{Bob},
$t(x,y)$ is a \emph{recipe} that an intruder can
use to replace $f(d_1,a)$ by $t(f(d_1,a),r(a'))$ to
generate a message compliant with the secret parameter $a$
currently used by \emph{Bob} to accept a message modified by \emph{Alice} by means of the secret parameter $a$. 
In this way the intruder can spoof \emph{Bob} into believing that $t(f(d_1,a),r(a'))$ came 
from \emph{Alice}. The function $r$ ensures that the 
randomly chosen parameter $r(a')$ meets condition (1), since 
otherwise there would be no spoofing. 
\noindent Let us see two examples of malleable lingos.

\begin{example} \label{ex:xor-mall}
The lingo $\Lambda_{xor}\{n\}$ from
Example \ref{ex:xor-lingo} is \textit{malleable}.  The recipe $t(x,y)$
 is $x \oplus y$  and $r_{\oplus}$ is the function
 $r_{\oplus} = \lambda y \in \{0,1\}^{n}.\; \mathbf{if}\; y \; .\!=\!. \;
 \vec{0} \;
 \mathbf{then} \; \vec{1} \; \mathbf{else} \; y \; \mathbf{fi}
 \in \{0,1\}^{n}$, where $.\!=\!.$ denotes the equality predicate
 on bitvectors.
 Therefore, $t(f(d_1,a),r(a'))=d_{1} \oplus a \oplus r(a')$,
 which meets condition $(1)$ because $r(a') \not= \vec{0}$
 and obviously meets condition $(2)$, since
 $t(f(d_1,a),r(a'))=f(d_1 \oplus r(a'),a)$.
\end{example}

\noindent An $f$-checkable lingo
provides a first line of defense against an intruder trying to
generate a payload compliant with a secret parameter $a$.
However, the example below shows that an $f$-checkable lingo can be malleable.

\begin{example} \label{ex:xor-sharp-mall}
The lingo $\Lambda_{\mathit{xor}}^{\sharp}\{n\}$
from Example \ref{ex:edxor} with $n \geq 2$
is comp-malleable with recipe\footnote{We assume without loss of
generality that $\Sigma_{\Lambda_{\mathit{xor}}^{\sharp}}$
contains projection functions $p_1,p_2 : D_2 \; D_2 \rightarrow D_2$ 
and $p_1,p_2 : A \; A \rightarrow A$ equationally specified by
the usual projection equations.}
$t(x,y)= (p_{1}(x) \oplus p_1(y),p_{2}(x) \oplus p_1(y))$
and function $r$ defined by:
\[
\lambda y \in \{0,1\}^{n} \otimes \{0,1\}^{n}.\;
\; \mathbf{if}\; \neg(p_1(y) {.=.} \vec{0}) \;
\mathbf{then} \; y \; \mathbf{else}\;
\]
\[\mathbf{if} \;
\neg(p_2(y) \; .\!=\!. \; \vec{1})\; \mathbf{then} \; (\vec{1},p_2(y)) \;
\mathbf{else} \; (0 \, \vec{1}_{n-1},p_2(y)) \; \mathbf{fi}\;
\]
\[
\mathbf{fi}
\in \{0,1\}^{n} \otimes \{0,1\}^{n},
\]
where
$0 \, \vec{1}_{n-1}$ denotes the $n$-bitvector starting with
$0$ and continued with $n-1$ $1$'s. Note that the key property of
$r(y)$ is that $p_1(r(y)) \not= \vec{0}$.  Therefore,
$t(f^{\sharp}(d_1,(a,a')),r((a'',a''')))=$
$(d_1 \oplus a \oplus p_1(r((a'',a'''))),
d_1 \oplus a' \oplus p_1(r((a'',a'''))))$
is in fact $f^{\sharp}(d_1 \oplus p_1(r((a'',a'''))) ,(a,a'))$,
which is different from $f^{\sharp}(d_1,(a,a'))
= (d_1 \oplus a,
d_1 \oplus a')$, thus
meeting conditions $(1)$ and $(2)$.
\end{example}

\noindent Examples \ref{ex:xor-mall}--\ref{ex:xor-sharp-mall}
suggest the following generalization:

\begin{theorem}\label{theorem:malleable} Let $\Lambda=(D_1,D_2,A,f,g)$ 
be a lingo where all data types and all functions are
computable, and such that for all $d_1 \in D_1$, $a,a',a'' \in A$, 
(i) $f(d_1,a) \in D_{1}$, (ii)
there is a computable
subset $A_{0} \subseteq A$ with at least two different elements
$a_0 , a_1$ and
such that
$a'' \in A_{0} \Rightarrow f(d_1,a'') \not= d_1$, and
(iii) $f(f(d_1,a),a')=f(f(d_1,a'),a)$.  Then both $\Lambda$
 and $\Lambda^{\sharp}$   are malleable.
\end{theorem}

\noindent {\bf Proof}:  To see that $\Lambda$ is malleable,
define $r = \lambda a \in A.\; {\bf if} \; a \in A_{0} \; {\bf then}
\; a \; {\bf else} \; a_{0} \; {\bf fi}$.  Then, recipe
$f(x,y)$ works, since by (ii), we have
$f(f(d_1,a),r(a')) \not= f(d_1,a)$,
and both $f(f(d_1,a),r(a'))$ 
and $f(f(d_1,r(a')),a),a)$ are compliant with $a$.

\vspace{1ex}

\noindent To see that $\Lambda^{\sharp}$ is malleable define
$r$ as the function:
\[r = \lambda y \in A \otimes A .\;
\; \mathbf{if}\; p_1(y) \in A_{0} \;
 \mathbf{then} \; y \; \mathbf{else}\;
 \]
 \[\mathbf{if} \;
 p_2(y) \; .\!=\!. \; a_{0} \; \mathbf{then} \;  (a_{1},p_2(y))  \;
 \mathbf{else} \; (a_{0},p_2(y)) \; \mathbf{fi}\;
  \]
  \[
  \mathbf{fi}
 \in A \otimes A.
 \]
$r$'s key property is of course that 
$\forall y \in A \otimes A, \; p_1 (r(y)) \in A_{0}$.  Then, recipe
$t(x,y)=[f(p_{1}(x),p_1(y)),f(p_{2}(x),p_1(y))]$ works
because
{\small
\[
t(f^{\sharp}(d_1,(a,a')),r((a'',a''')))=
\]
\[
[f(f(d_1,a),p_1(r((a'',a''')))),
 f(f(d_1,a'),p_1(r((a'',a'''))))] =
\]
\[
[f(f(d_1,p_1(r((a'',a''')))),a),
 f(f(d_1,p_1(r((a'',a'''))),a')]
\]
}
so that, by the definition of $r$ and (ii) we have,
\[t(f^{\sharp}(d_1,(a,a')),r((a'',a''')))\not=
f^{\sharp}(d_1,(a,a'),r((a'',a''')))
\]
and we of course have
$(f(f(d_1,p_1(r((a'',a''')))),a), \\
 f(f(d_1,p_1(r((a'',a''')))),a'))$ compliant with $(a,a')$,
as desired. $\Box$

\vspace{2ex}

\noindent Being malleable is an undesirable property.  One would 
like to transform
a  possibly malleable lingo $\Lambda$ into a 
non-malleable lingo $\Lambda'$.  Developing general methods for
proving that a lingo is non-malleable is a topic for future research. 
However, in 
Section~\ref{non-mall-remarks} 
we discuss several methods that we conjecture could support lingo transformations 
$\Lambda \mapsto \Lambda'$ yielding non-malleable lingos $\Lambda'$.

\subsection{Authenticating Lingos}\label{subsubsec:authenticating-lingos}

Here we extend our threat model to one in which it is possible that some enclave members are dishonest and may try to impersonate other enclave members.  In this case we  assume each pair of principals that communicate with each other shares a unique secret.  (When a client-server architecture is used, this means that the number of secrets is only equal to the number of clients.)  We show how a lingo whose secret parameter $a$ is updated every time it is used can be turned into a lightweight \emph{authenticating lingo} below.

The key idea about an authenticating lingo is that, if Bob receives
a transformed payload $f(d_1,a)$ supposedly from Alice, he has a way
to check that it was sent by Alice because he can extract from
$f(d_1,a)$ some data that was produced using
secret information  that only Alice and Bob share.  In this case it will be the result of computing a one-way hash function $\mathit{hash}$ over their identities and their current shared parameter $a$. To make this possible, we need to make explicit two
data types involved in the communication.  First, honest participants
in the communication protocol using lingo $\Lambda$ typically have 
corresponding \emph{object identifiers} that uniquely identify such 
participants and
belong to a finite data type $\mathit{Oid}$.
Second, the result of computing the $\mathit{hash}$  belongs to another finite data type $H$.  Here is
the definition:

\begin{definition} An \emph{authenticating lingo} is a tuple 
$((D_1 , D_2 ,A,f,g),\mathit{Oid},H,\mathit{param}, \\
\mathit{hash},\mathit{code})$
such that:
\begin{enumerate}
\item $(D_1 , D_2 ,A,f,g)$ is a lingo;

\item $\mathit{Oid}$ and $H$ are finite data types of, respectively,
object identifiers and outputs of $\mathit{hash}$.

\item $\mathit{param}$ is a function $\mathit{param}: \mathbb{N} \times  \mathit{Oid} \otimes \mathit{Oid} \rightarrow A$;

\item $\mathit{hash}$ is a function $\mathit{hash}: \mathbb{N} \times  \mathit{Oid} \otimes \mathit{Oid} \rightarrow H$;\footnote{$\mathit{hash}$ should have good
properties as a hash function.  This seems impossible, since $H$ is finite
but $\mathbb{N}$ is not.  But this problem is only apparent.  In practice the
values $n \in \mathbb{N}$ used as first arguments of $\mathit{hash}$
(and also as first arguments of $\mathit{param}$)
will be random numbers $n$ such that $n \in \mathbb{N}_{< 2^{k}}$
for some fixed $k$.} and

\item $\mathit{code}$ is a function $\mathit{code}: D_2 \times A \rightarrow H$
\end{enumerate}
such that $\forall d_1 \in D1$, $\forall n \in \mathbb{N}$, $\forall (A,B) \in \mathit{Oid} \otimes \mathit{Oid}$,
\[\mathit{code}(f(d_1,\mathit{param}(n,(A,B))),\mathit{param}(n,(A,B))) = 
\mathit{hash}(n,(A,B)).\]
\end{definition}

\noindent Let us illustrate this notion by means of a transformation
$\Lambda \mapsto \Lambda^{\alpha}$ that generates an authenticating
lingo $\Lambda^{\alpha}$ from a lingo $\Lambda$.  For concreteness,
assume that $\Lambda$ is of the form $\Lambda = (\mathbb{N}_{< 2^{n}},
\mathbb{N}_{< 2^{m}},A_{0},f_{0},g_{0})$ with $A_{0}$ a finite set,
and that there is a function $\mathit{param}_{0}: \mathbb{N}_{\leq 2^{k}}
\rightarrow A_{0}$ assigning to each random number $n \in \mathbb{N}_{< 2^{k}}$
a corresponding parameter in $A_{0}$.  This reflects the fact that, in practice,
two honest participants will share a common parameter $a_{0} \in A_{0}$
by sharing a common secret random number $n$ and using $\mathit{param}_{0}$
to generate $a_0$.
Assuming a finite data type $\mathit{Oid}$ of honest participants, 
choose a function $\mathit{hash}: \mathbb{N}_{< 2^{k}} \times  \mathit{Oid} \otimes \mathit{Oid} \rightarrow \mathbb{N}_{< 2^{j}}$,
with $j$ and $k$ big enough for $\mathit{hash}$
to be computationally infeasible to invert.
Choose a function $\iota: \mathbb{N}_{< 2^{k}} \rightarrow \mathit{Invo}(m+j)$ mapping each random number $n$ to an \emph{involution}\footnote{An involution
is a permutation $\sigma$ such that $\sigma$ is its own inverse permutation, i.e.,
$\sigma = \sigma^{-1}$.} permutation $\sigma$
on $m+j$ elements. 
By abuse of notation, if $\vec{b}= b_1 , \ldots , b_{m + j}$ is a bit sequence of length $m + j$,
we let $\sigma(b_1 , \ldots , b_{t}) = b_{\sigma(1)} , \ldots , b_{\sigma(m+j)}$.

Then define $\Lambda^{\alpha}$ as follows:
$
\Lambda^{\alpha} = ((\mathbb{N}_{< 2^{n}},
\mathbb{N}_{< 2^{m+j}},A_{0} \times \mathit{Invo}(m+j) \times \mathbb{N}_{< 2^{j}} ,f,g), \\
\mathit{Oid},\mathbb{N}_{< 2^{j}},\mathit{param},\mathit{hash},\mathit{code})$
where:
\begin{enumerate}
\item $f(d_1 ,(a_{0}, \sigma,d)) = \sigma(f_{0}(d_1 ,a_0 );d)$,
where we view $f_{0}(d_1 ,a_0 )$ and $d$ as bit vectors of
respective lengths $m$ and $j$, $f_{0}(d_1 ,a_0 );d$ denotes
their concatenation as bit sequences. 

\item $g(\vec{b},(a_{0}, \sigma,d))=g_{0}(p_{1}(\sigma(\vec{b})),a_{0})$,
where by abuse of notation identify $\vec{b}$ with a number smaller than $2^{m+j}$, and where $p_1$
(resp. $p_{2}$)
projects an $m+j$ bitvector to its first $m$ bits (resp. its last $j$ bits).

\item $\mathit{param}: \mathbb{N}_{< 2^{k}} \times  \mathit{Oid} \otimes \mathit{Oid} \rightarrow A_{0} \times \mathit{Invo}(m+j) \times \mathbb{N}_{< 2^{j}}$ is the function
$\lambda (n,(A,B)).\; (param_{0}(n),\iota(n),\mathit{hash}(n,(A,B)))$.

\item $\mathit{code}(\vec{b},(a_{0}, \sigma,d))= p_{2}(\sigma(\vec{b}))$.
\end{enumerate}

The reader can check that $\Lambda^{\alpha}$ is an authenticating 
lingo.

\section{Lingo Compositions} \label{sec:compositions}

\noindent Lingos can be composed in various ways to obtain new lingos. Lingo compositions provide modular, automated methods of obtaining new lingos from existing ones.  Such compositions
are often stronger, i.e., harder to compromise by an attacker, than the lingos so composed. We define four such lingo composition operations, namely, \emph{horizontal} and \emph{functional} composition of lingos,
\emph{products} of lingos, and \emph{tuplings} of lingos.
We also explain how a lingo's input and output data types can be 
modified by means of \emph{data-adaptors}.

\subsection{Horizontal Composition of Lingos}\label{subsubsec:horizontal-composition}

\noindent Given a finite family of lingos $\vec{\Lambda} = \{\Lambda_{i}\}_{1 \leq i \leq k}$, $k \geq 2$, all sharing the same input data type $D_1$,
their \emph{horizontal composition}, denoted $\bigoplus \vec{\Lambda}$
is intuitively
their union, i.e., in $\bigoplus \vec{\Lambda}$ the input data type
$D_1$ remains the same,
$D_2$ is the union of the $D_{2.i}$, $1 \leq i \leq k$,
and $A$ is the disjoint union of the $A_{i}$, $1 \leq i \leq k$.
$\bigoplus \vec{\Lambda}$ is a more complex lingo that each of its 
component lingos
$\Lambda_{i}$, making it harder to compromise
by an attacker, because it becomes a hydra with $k$ heads:
quite unpredictably,
sometimes behaves like some $\Lambda_{i}$ 
and sometimes like another $\Lambda_{j}$.  Furthermore,
it has a bigger parameter space, since
$|A| = \Sigma_{1 \leq i \leq k} |A_{i}|$.

\begin{definition}[Horizontal Composition]
Let $\vec{\Lambda}$ be a finite family of lingos
$\vec{\Lambda} = \{\Lambda_{i}\}_{1 \leq i \leq k}$,
$k \geq 2$, all having the same input data type $D_1$, i.e.,
$\Lambda_{i}=(D_{1},D_{2.i},A_{i},f_{i},g_{i})$, $1 \leq i \leq k$.
Let $\vec{d}_{0}=(d_{0.1},\ldots,d_{0.k})$ be a choice of default
$D_{2.i}$-values, $d_{0.i}\in D_{2.i}$, $1 \leq i \leq k$.
Then, the \emph{horizontal composition} of the lingos
$\vec{\Lambda}$ with default $D_{2.i}$-values  $\vec{d}_{0}$
is the lingo:
\[
\bigoplus_{\vec{d_{0}}} 
\vec{\Lambda}=(D_{1}, \bigcup_{1 \leq i \leq k} D_{2.i},
\bigcup_{1 \leq i \leq k} A_{i} \times\{i\}
,\oplus \vec{f},\oplus \vec{g})
\]
where, for each $d_1 \in D_1$, $d_2 \in \bigcup_{1 \leq i \leq k} D_{2.i}$, and $a_{i} \in A_{i}$, $1 \leq i \leq k$,
\begin{enumerate}
\item $\oplus \vec{f} = \lambda (d_1,(a_i ,i)).f_i(d_1,a_i)$,
\item $\oplus \vec{g} = \lambda (d_2 ,(a_i ,i)).\; 
\mathbf{if} \; d_{2} \in D_{2.i} \; \mathbf{then} \; g_{i}(d_i ,a_i) \; \mathbf{else} \\ \; g_{i}(d_{0.i},a_{i})$.
\end{enumerate}
$\bigoplus_{\vec{d_{0}}} \vec{\Lambda}$ is indeed a lingo:
$\oplus \vec{g}(\oplus \vec{f}(d_1,(a_{i},i)),(a_i , i))=
g_{i}(f_{i}(d_1,a_i ),a_i )= d_1$.
%

\vspace{2ex}

\noindent We have already pointed out that, in practice, two
protocol participants using a lingo to first modify and then decode a payload $d_1$ that, say, Alice sends to Bob, agree on a secret parameter
$a \in A$ by agreeing on a secret random number $n$,
because both use a function $\mathit{param}: \mathbb{N} \rightarrow A$
to get $a = \mathit{param}(n)$.  This then poses the practical
problem of synthesizing a function $\oplus \vec{\mathit{param}}: \mathbb{N} \rightarrow \bigcup_{1 \leq i \leq k} A_{i} \times \{i\}$
for $\bigoplus_{\vec{d_{0}}} \vec{\Lambda}$
out of the family a functions $\{ \mathit{param}_{i} : \mathbb{N} \rightarrow A_{i}\}_{1 \leq i \leq k}$ used in each of the lingos $\Lambda_{i}$.
Furthermore, different lingos in the family $\vec{\Lambda}$
may have different degrees of strength against an adversary.
This suggest favouring the choice of stronger lingos over that
of weaker lingos in the family $\vec{\Lambda}$ to achieve a
function $\oplus \vec{\mathit{param}}$ that improves the overall strength of 
$\bigoplus_{\vec{d_{0}}} 
\vec{\Lambda}$.  This can be achieved by choosing a \emph{bias vector}
$\vec{\beta}=(\beta_1,\ldots,\beta_k) \in \mathbb{N}_{>0}^{k}$,
so that, say, if lingo $\Lambda_i$ is deemed to be stronger than lingo $\Lambda_j$, then 
the user chooses $\beta_i > \beta_j$.  That is, $\vec{\beta}$
specifies a \emph{biased dice} with $k \geq 2$ faces,
so that the dice will show face $j$ with probability $\frac{\beta_{j}}{\Sigma_{1 \leq i \leq k} \beta_{i}}$.  Therefore, we can use a
pseudo-random function $\mathit{throw}_{\vec{\beta}}: \mathbb{N} \rightarrow \{1,\ldots, k\}$ simulating a sequence of throws of a $k$-face dice with bias $\vec{\beta}$
to get our desired function 
$\oplus \vec{\mathit{param}}$ as the function:
\[
\oplus \vec{\mathit{param}}(n) = (\mathit{param}_{\mathit{throw}_{\vec{\beta}}(n)}(n),\mathit{throw}_{\vec{\beta}}(n)).
\]
\end{definition}

\begin{example} \label{ex:dccomp} To see how horizontal composition can strengthen lingos, consider the D\&C lingo described in Example \ref{ex:d&c-lingo-old}, in which $ f(n,a) = (x,y)$  where $x = quot(n +(a+2),a +2)$ and $y = rem(n + (a+2),a+2)$. We note that any choice of $0 \le y < a + 2$ will pass the $f$-check, so the choice of $y = 0$ or $1$ will always pass the $f$-check. To counter this,   let \emph{reverse} D\&C be the  lingo with $f(n,a) = (y,x)$, where $y$ and $x$ are computed as in D\&C.  The attacker's best strategy in reverse  D\&C is the opposite of that in D\&C.  Thus composing D\&C  and reverse D\&C  horizontally with a bias vector $(.5,.5)$ means that the attacker strategy of choosing the first (respectively, second) element of the output to be 0 succeeds with probability 0.5 in any particular instance, as opposed to probability 1 for D\&C by itself.
\end{example}

\noindent 
The following result is immediate.

\begin{lemma}
If each $\Lambda_i$ in $\vec{\Lambda} = \{\Lambda_{i}\}_{1 \leq i \leq k}$,
$k \geq 2$, is $f$-checkable, then $\bigoplus_{\vec{d_{0}}} 
\vec{\Lambda}$ 
is also $f$-checkable.
\end{lemma}

\noindent The horizontal composition of $k$ lingos
sharing a common input data type
can be specified in Maude as a functional module parametrised 
by the $k$-\texttt{LINGO} parameter theory.  We illustrate 
here the construction for $k=2$, where $2$-\texttt{LINGO}
is abbreviated to \texttt{DILINGO}.  Likewise
for $k=3$, where $3$-\texttt{LINGO}
is abbreviated to \texttt{TRILINGO}.

\begin{Verbatim}
fth DILINGO is
    sorts D1 D21 D22 A1 A2 .
    
    op f1 : D1 A1 -> D21 . op f2 : D1 A2 -> D22 .
    op g1 : D21 A1 -> D1 . op g2 : D22 A2 -> D1 .
    op d0.1 : -> D21 . op d0.2 : -> D22 .

    var a1 : A1 . var a2 : A2 . var d1 : D1 .
    eq g1(f1(d1, a1), a1) = d1 .
    eq g2(f2(d1, a2), a2) = d1 .
endfth

fmod HOR-COMP{DL :: DILINGO} is
    sorts A D2 . subsorts DL$D21 DL$D22 < D2 .
    
    op [_]1 : DL$A1 -> A [ctor] . op [_]2 : DL$A2 -> A [ctor] .
    
    op +f : DL$D1 A -> D2 . op +g : D2 A -> DL$D1 .
    
    var d1 : DL$D1 .  var a1 : DL$A1 . var a2 : DL$A2 . 
    var d21 : DL$D21 . var d22 : DL$D22 .  var d2 : D2 .
    
    eq +f(d1,[a1]1) = f1(d1,a1) . eq +f(d1,[a2]2) = f2(d1,a2) .
    eq +g(d21,[a1]1) = g1(d21,a1) . eq +g(d22,[a2]2) = g2(d22,a2) .
    ceq +g(d2,[a1]1) = g1(d0.1,a1) if d2 :: DL$D21 = false .
    ceq +g(d2,[a2]2) = g2(d0.2,a2) if d2 :: DL$D22 = false .
endfm
\end{Verbatim}

\begin{example}[Horizontal composition of XOR-BSeq and D\&C Lingos]
\label{ex:horizontal-comp}
The lingos $\Lambda_{xor.\mathit{BSeq}}$
of Example \ref{ex:XOR-BSeq-Lingo}
and $\lingo_{D\&C}$ of Example \ref{ex:d&c-lingo-old}
share the same $D_{1}$, namely, $\mathbb{N}$.
They do therefore have a horizontal composition
$\Lambda_{xor.\mathit{BSeq}} \oplus_{\vec{d}_{0}} \lingo_{D\&C}$
for any choice of $\vec{d}_{0}$.  We leave as an exercise for the
reader to define an appropriate view \texttt{v} from
the theory \verb@DLINGO@ to \verb@NAT-PAIR@ in Maude,
so that the horizontal composition $\Lambda_{xor.\mathit{BSeq}} \oplus_{\vec{d}_{0}} \lingo_{D\&C}$ is just the instantiation
\verb@HOR-COMP{v}@.
\end{example}

\subsection{Functional Composition of Lingos}\label{subsubsec:functional-composition}

Given two lingos $\Lambda$ and $\Lambda'$ such that the output data type
$D_2$  of $\Lambda$ coincides with the input data type of $\Lambda'$,
it is possible to define a new lingo $\Lambda \odot \Lambda'$
whose $f$ and $g$ functions are naturally the compositions
of $f$ and $f'$ (resp. $g$ and $g'$) in a suitable way.
$\Lambda \odot \Lambda'$ is clearly more complex and harder to
compromise than either $\Lambda$ or $\Lambda'$.  In particular,
its parameter space is  $A \times A$ with cardinality
$|A| \times |A'|$; a much bigger set whose secret parameters
$(a,a')$ are considerably harder to guess than either $a$ or $a'$.

\begin{definition}[Functional Composition]
Given lingos $\Lambda=(D_1,D_2,A,f,g)$ and $\Lambda'=(D_2,D_3,A',f',g')$, 
their \emph{functional composition} is the lingo
$\Lambda \odot \Lambda' =
(D_1,D_3,A \times A',f \cdot f',g * g')$, where
for each $d_1 \in D_1$, $d_3 \in D_3$, and 
$(a,a') \in A \times A'$,
\begin{itemize}
\item $f \cdot f'(d_1,(a,a')) =_{\mathit{def}}
f'(f(d_1,a),a')$,
\item $g * g'(d_3,(a,a')) =_{\mathit{def}}
g(g'(d_3,a'),a)$.
\end{itemize}
$(D_1,D_3,A \times A',f \cdot f',g * g')$ is indeed a lingo,
since we have:
\begin{align*}
    g * g'(f \cdot f'(d_1,(a,a')),(a,a')) &= g(g'(f'(f(d_1,a),a'),a'),a) \\
    &=g(f(d_1,a),a) = d_1.
\end{align*}
\end{definition}

\noindent The  functional composition
transformation $(\Lambda,\Lambda') \mapsto \Lambda \odot \Lambda'$
can be formally specified in Maude
with a parametrised module, parametric
on the 
theory \texttt{COMP-LINGOS} 
of (functionally) composable
lingos. 

\begin{Verbatim}
fth COMP-LINGOS is
    sorts D1 D2 D3 A A'.
    op f : D1 A -> D2 . op f' : D2 A' -> D3 .
    op g : D2 A -> D1 . op g' : D3 A' -> D2 .
    
    var d1 : D1 . var d2 : D2 . var a : A . var a' : A' .
    eq g(f(d1,a),a) = d1 . eq g'(f'(d2,a')) = d2 .

    op param : Nat -> A . op param' : Nat -> A' .
endfth

fmod FUN-COMP{CL :: COMP-LINGOS} is
    sort pairA . op a[_,_] : CL$A CL$A' -> pairA [ctor] .
    
    var d1 : D1 . var d3 : D3 . var a : A . var a' : A' . var n : Nat .
    op f.f' : D1 pairA -> D3 . eq f.f'(d1,a[a,a']) = f'(f(d1,a),a') .
    op g*g' : D3 pairA -> D1 . eq g*g'(d3,a[a,a']) = g(g'(d3,a'),a) .

    op param.param' : Nat -> pairA . 
    eq param.param'(n) = a[param(n),param'(n)] .
endfm
\end{Verbatim}

\begin{example}[Functional composition of XOR-BSeq and D\&C Lingos]
\label{ex:functional-comp}
The lingos $\Lambda_{xor.\mathit{BSeq}}$
of Example \ref{ex:XOR-BSeq-Lingo}
and $\lingo_{D\&C}$ of Example \ref{ex:d&c-lingo-old}
are functionally composable as
$\Lambda_{xor.\mathit{BSeq}} \odot \lingo_{D\&C}$,
because $\mathbb{N}$ is both the output
data type of $\Lambda_{xor.\mathit{BSeq}}$ and
the input data type of $\lingo_{D\&C}$.
It is possible to define an appropriate view \texttt{w} from
the theory \verb@COMP-LINGOS@ to \verb@NAT-PAIR@ in Maude,
so that the functional composition 
$\Lambda_{xor.\mathit{BSeq}} \odot \lingo_{D\&C}$
is just the instantiation
\verb@FUN-COMP{w}@.  Note that, by Lemma \ref{f-check-4-fun-lemma}
below, and Theorem \ref{theorem:trasnform-fCheckable},
$\Lambda_{xor.\mathit{BSeq}} \odot \lingo_{D\&C}$
is $f$ checkable, in spite of $\Lambda_{xor.\mathit{BSeq}}$
not being so.
\end{example}

\begin{lemma}\label{f-check-4-fun-lemma}
Given lingos $\Lambda=(D_1,D_2,A,f,g)$ and $\Lambda'=(D_2,D_3,A',f',g')$, if $\Lambda'$ if $f$-checkable,
then the \emph{functional composition} 
$\Lambda \odot \Lambda'$ is also $f$-checkable.
\end{lemma}

\vspace{2ex}

\noindent {\bf Proof}: We need to show that for each $(a,a') \in A \times A'$
there exists $d_3 \in D_3$ such that $\nexists d_1 \in D1$ such that
$d_3 = f'(f(d_1,a),a')$.  But, by assumption, there exists
a $d_3 \in D_3$ such that $\nexists d_2 \in D2$ such that
$d_3 = f'(d_2,a')$.  This fact applies, in particular,
to $d_2 = f(d_1,a)$.  $\Box$

\vspace{2ex}

\subsection{Modifying and Composing Lingos by means of Data Adaptors}\label{subsubsec:data-adaptor-composition}

Two lingos $\Lambda=(D_1,D_2,A,f,g)$ and 
$\Lambda'=(D_2',D_3,A',f',g')$ will not be composable via functional
composition if $D_2 \not= D'_2$. However, they may be
functionally composable by using a \emph{data adaptor} that ``connects''
the data types $D_2$ and $D'_2$ (i.e makes them equal).  The most 
obvious case is when $D_2$ and $D'_2$ are in bijective 
correspondence, i.e., when, as data types, they are identical
up to a change of data representation. For example,
$D_2$ may be a binary representation of naturals and $D'_2$
a decimal representation of naturals. This can be
generalized to the case when
$D_2$ is in bijective correspondence with a subset of $D'_2$
by means of a \emph{section-retract pair} $(j,r)$ of functions
$j: D_2 \rightarrow D'_2$  and $r: D_2' \rightarrow D_2$
such that $\forall d_2 \in D_2,\; r(j(d_2))=d_2$. 
A section-retract pair $(j,r)$ will be denoted
$(j,r) : D_2 \rightleftarrows D'_2$.  Note that
this forces $j$ to be injective and $r$ surjective, and therefore
defines a bijective correspondence between 
$j(D_2)$ and $D'_2$. Note also that
$(j,r)$ and $(r,j)$ are both section-retract pairs iff
$j: D_2 \rightarrow D'_2$ is bijective with inverse
bijection $j^{-1}=r$.  Therefore, section-retract pairs
generalize bijective correspondences.  We will treat 
the general section-retract pair case, leaving the bijective 
special case to the reader. Furthermore, in 
what follows we will call a section-retract pair
$(j,r) : D_2 \rightleftarrows D'_2$ a \emph{data adaptor}.

The first key observation
is that, given a lingo $\Lambda=(D_1,D_2,A,f,g)$
we can obtain another lingo by either adapting $D_1$
by means of a data adaptor $(j,r) : D'_1 \rightleftarrows D_1$,
or adapting $D_2$ by means of a data adaptor 
$(j',r') : D_2 \rightleftarrows D'_2$.

\begin{definition}\label{def:adapted-lingos}
Given a lingo $\Lambda=(D_1,D_2,A,f,g)$
and data adaptors $(j,r) : D'_1 \rightleftarrows D_1$
and $(j',r') : D_2 \rightleftarrows D'_2$, 
we define
$(j,r); \Lambda$ and $\Lambda ; (j',r')$
as follows:
\begin{itemize}
\item $(j,r); \Lambda = (D'_1,D_2,A,\widehat{f},\widehat{g})$, 
s.t. $\forall d'_1 \in D'_1,\; \forall d_2 \in D_2, \; \forall a \in A,$
    $\widehat{f}(d'_1,a)=f(j(d'_1),a)$ and 
    $\; \widehat{g}(d_2,a)=r(g(d_2,a))$

\item $\Lambda ; (j',r') = (D_1,D'_2,A,\widetilde{f},\widetilde{g})$, 
s.t. $\forall d_1 \in D'_1,\; \forall d'_2 \in D'_2, \; \forall a \in A,$
    $\widetilde{f}(d_1,a)=j'(f(d_1,a))$ and 
    $\; \widetilde{g}(d'_2,a)=g(r'(d'_2),a)$
\end{itemize}
\end{definition}

\begin{theorem}\label{theorem:adapted-lingos-are-lingos} $(j,r); \Lambda$ and $\Lambda ; (j',r')$
from Definition~\ref{def:adapted-lingos} are both lingos.
\end{theorem}

\noindent {\bf Proof}: 
$(j,r); \Lambda = (D'_1,D_2,A,\widehat{f},\widehat{g})$ is a lingo because,
\[
\widehat{g}(\widehat{f}(d'_1,a),a)=r(g(f(j(d'_1),a),a))=
r(j(d'_1))=d'_1
\]

\noindent Likewise, $\Lambda ; (j',r') = (D_1,D'_2,A,\widetilde{f},\widetilde{g})$ is a lingo because, 
\[
\widetilde{g}(\widetilde{f}(d_1,a),a)=
g(r'(j'(f(d_1,a)),a))=g(f(d_1,a),a)=d_1
\]
$\Box$

\begin{theorem}\label{theorem:adapted-lingos-equality} For $(j,r); \Lambda$ and $\Lambda ; (j',r')$
from Definition~\ref{def:adapted-lingos} the following equality of lingos holds:
\[
((j,r); \Lambda);(j',r')=(j,r);(\Lambda ; (j',r'))
\]
\end{theorem}

\noindent {\bf Proof}:  Let use denote
$((j,r); \Lambda);(j',r')=(D'_1,D'_2,A,f_1,g_1)$
and $(j,r);(\Lambda ; (j',r'))= (D'_1,D'_2,A,f_2,g_2)$.
We need to show that $f_1=f_2$ and $g_1=g_2$.
Indeed, for all $d'_1 \in D'_1$, $d_2 \in D'_2$
and $a \in A$ we have:

\[
f_1(d'_1,a)= j'(\widehat{f}(d'_1,a))=j'(f(j(d'_1),a))=
\widetilde{f}(j(d'_1),a)= f_2(d'_1,a)
\]
\[
g_1(d'_2,a)=\widehat{g}(r'(d'_2,a))=r(g(r'(d'_2,a)))=
r(\widetilde{g}(d'_2,a))=g_2(d'_2,a)
\]
$\Box$

The following result shows how data adaptors can enable
functional compositions of lingos that could not be so composed otherwise.

\begin{theorem}\label{theorem:adapted-lingos-functional} Let $\Lambda=(D_1,D_2,A,f,g)$ and 
$\Lambda'=(D_2',D_3,A',f',g')$ be lingos,
$(j,r) : D_2 \rightleftarrows D'_2$ a data adaptor, and
all the above-mentioned data types and functions computable.  Then,
the following two functional compositions of lingos are
identical:
\[(\Lambda ; (j,r)) \odot \Lambda'
= \Lambda \odot ((j,r) ; \Lambda')
\]   
\end{theorem}

\noindent {\bf Proof}:
Let $(\Lambda ; (j,r)) \odot \Lambda'=
(D_1,D_3,A \times A',f_1,g_1)$, and
$\Lambda \odot ((j,r) ; \Lambda')= 
(D_1,D_3,A \times A',f_2,g_1)$.  We need to
show that $f_1=f_2$ and $g_1=g_2$.
Indeed, for all $d_1 \in D_1$, $d_3 \in D_3$
and $(a,a') \in A \times A'$ we have:
\begin{align*}
f_1(d_1,(a,a'))&= f'(\widetilde{f}(d_1,a),a')=f'(j(f(d_1,a)),a')\\
&=\widehat{f'}(f(d_1,a),a')=f_2(d_1,(a,a')),
\end{align*}
\begin{align*}
g_1(d_3,(a,a'))&=\widetilde{g}(g'(d_3,a'),a)=g(r(g'(d_3,a')),a)\\
&=g(\widehat{g'}(d_3,a'),a)=g_2(d_3,(a,a')).
\end{align*}
$\Box$

\subsection{Cartesian Products and Tupling of Lingos}

\noindent Recall that a lingo is just a (non-trivial) 3-sorted $(\Sigma,E)$-algebra.
By Birkhoff's Variety Theorem, $(\Sigma,E)$-algebras are closed
under Cartesian products.  Since products of non-trivial
$(\Sigma,E)$-algebras are non-trivial, this means that
lingos are closed under Cartesian products. The simplest
case is that of a binary Cartesian product. Given
any two lingos $\Lambda=(D_1,D_2,A,f,g)$ and 
$\Lambda'=(D_1',D'_2,A',f',g')$, its \emph{Cartesian product}
$\Lambda \times \Lambda'$
is the lingo,
$\Lambda \times \Lambda'=
(D_1\times D'_1,D_2 \times D'_2,A \times A',f \times f',g \times g')$
where, by definition, 
\[
f \times f'= \lambda ((d_1,d'_1),(a.a')).\;(f(d_1,a),f'(d'_1,a')) \text{, and}
\]
\[
g \times g'= \lambda ((d_2,d'_2),(a.a')).\;(g(d_2,a),g'(d'_2,a')).
\]
This generalizes in a straightforward way to the product of
$k$ lingos, $k \geq 2$, i.e.,
Given a family of $k$ lingos
$\{\Lambda_{i} =(D_{1.i},D_{2.i},A_i,f_i,g_i) \}_{1 \leq i \leq k}$,
their \emph{Cartesian product} $\Lambda_{1} \times \ldots \times\Lambda_{k}$
is the lingo:
{\small
$\Lambda_{1} \times \ldots \times\Lambda_{k}=
(D_{1.1}\times \ldots \times D_{1.k},D_{2.1}\times \ldots \times D_{2.k},
A_{1}\times \ldots \times A_{k},f_{1}\times \ldots \times f_{k},
g_{1}\times \ldots \times g_{k}).$
}

\noindent A related lingo composition operation
is the tupling of lingos which share the same input data type.
Given lingos $\Lambda=(D_1,D_2,A,f,g)$ and 
$\Lambda'=(D_1,D'_2,A',f',\\g')$, their \emph{tuping}
$[\Lambda,\Lambda']$ is the 5-tuple,
$[\Lambda,\Lambda']=(D_1,D_2 \times D'_2,A \times A',[f,f'],g \lhd g')$
where, by definition, $[f,f']= \lambda (d_1,(a,a')).\; (f(d_1,a),f'(d_1,a'))$
and $g \lhd g'= \lambda ((d_2,d'_2),(a,a')).\; g(d_2,a)$.
$[\Lambda,\Lambda']$ is indeed a lingo, since we have:
\begin{align*}
    (g \lhd g')([f,f'](d_1,(a,a')) &= (g \lhd g')(f(d_1,a),f'(d_1,a')) 
    =g(f(d_1,a),a) = d_1.
\end{align*}
Tupling generalizes in a straightforward way to the product of
$k$ lingos, $k \geq 2$ sharing a common input data type, i.e.,
given a family of $k$ lingos
$\{\Lambda_{i} =(D_1,D_{2.i},A_i,f_i,g_i) \}_{1 \leq i \leq k}$,
their \emph{tupling} $[\Lambda_{1} \times \ldots \times\Lambda_{k}]$
is the lingo:
\begin{align*}
    [\Lambda_{1},\ldots,\Lambda_{k}]=(&D_{1},D_{2.1}\times \ldots \times D_{2.k}, 
    A_{1}\times \ldots \times A_{k},[f_{1},\ldots,f_{k}], 
    g_{1}\lhd g_{2}\ldots ,g_{k})
\end{align*}
with the obvious definition of $[f_{1},\ldots,f_{k}]$
and $g_{1}\lhd g_{2}\ldots ,g_{k}$ generalizing the case $k=2$.

\vspace{2ex}

\noindent To illustrate the usefulness of the tupling operation
we need yet another simple remark: by Birkhoff's Variety
Theorem, $(\Sigma,E)$-algebras are closed under \emph{subalgebras}.
In particular, if 
$\Lambda=(D_1,D_2,A,f,g)$ is a lingo, a \emph{sublingo}
is a $\Sigma$-algebra $\Lambda'=(D'_1,D'_2,A',f',g')$
such that $D'_1 \subseteq D_1$, $D'_2 \subseteq D_2$,
$A' \subseteq A$, $f' \subseteq f$ and $g' \subseteq g$.
Since we wish $\Lambda'$ to be non-trivial, we also require that
$D'_1,D'_2$ and $A'$ are all non-empty.
$\Lambda'$ is then a lingo because $\Lambda$ is so,
as Birkhoff pointed out.  We then write
$\Lambda' \subseteq \Lambda$ to denote the sub-lingo relation.
Here is an interesting example:

\begin{example} Let $\Lambda=(D_1,D_2,A,f,g)$ be a lingo.
Recall its $f$-checkable transformed lingo $\Lambda^{\sharp}$
from Theorem \ref{theorem:trasnform-fCheckable}.  Note that
we have a sub-lingo inclusion,
$\Lambda^{\sharp} \subseteq [\Lambda,\Lambda].$
\end{example}

\subsection{Towards Non-malleable Lingos} \label{non-mall-remarks}

\noindent In hindsight, lingo transformations and lingo
compositions may provide useful methods to
obtain non-malleable lingos.  We discuss below some
methods that we conjecture will be useful for this purpose.
A first method
is applying a
transformation of the form $\Lambda \mapsto (j,r); \Lambda$,
where $(j,r) : D_0 \rightleftarrows D_1$
is such that $j(D_0) \subset D_1$ is
a ``sparse'' subset of $D_1$.  For example, $\Lambda$
can be $\Lambda_{\mathit{xor}}$ or $\Lambda_{\mathit{xor}}^{\sharp}$
and $D_0$ can be the set of all the words in the English language, 
with $n$ large enough so that $j$ can represent words 
as bit-vectors
in $\{0,1\}^{n}$ by mapping each English word to a (possibly padded)
concatenation of the ASCII bit-vectors for its letters.  
%
The intruder will see an obfuscated message of the
form $j(d_0) \oplus a$ with $d_0$ a word in English,
and using the recipe $x \oplus y$
will replace it by the obfuscated message
$j(d_0) \oplus a \oplus r_{xor}(a')$.
But recipe $x \oplus y$ is unlikely to pass muster
for $(j,r); \Lambda_{\mathit{xor}}$,
since, due to the sparsity of $j(D_0)$ in $D_1$,
it is highly unlikely that the $f$-checking equality test
$j(d_0) \oplus r_{xor}(a') =j(r(j(d_0) \oplus r_{xor}(a'))$
will hold, since this would exactly mean that 
$j(d_0) \oplus r_{xor}(a')$ is (the ASCII bit-vector representation
of) a word in English.  This is not a proof
that $(j,r); \Lambda_{\mathit{xor}}$ is non-malleable:
it is just a hint that it seems unlikely to be malleable.
A similar hint could be given for
$(j,r) ;\Lambda_{\mathit{xor}}^{\sharp}$.

Given a malleable lingo $\Lambda$, we conjecture
that a second method to transform it into a non-malleable
one may be by means of a functional composition 
$\Lambda  \odot \Lambda'$
with a suitably chosen lingo $\Lambda'$.  For example,
$\Lambda$ may satisfy the equation $f(f(d_{1},a_1),a_2)=f(f(d_{1},a_2),a_1)$.
But $\Lambda'$ may be chosen so that the
corresponding equation for $\Lambda  \odot \Lambda'$, namely,
\[
f'(f(f'(f(d_1 ,a_1),a'_1),a_2),a'_2) =
f'(f(f'(f(d_1 ,a_2),a'_2),a_1),a'_1),
\]
where $a_1, a_2 \in A$ and $a'_1, a'_2 \in A'$, fails to hold,
thwarting the success of some recipes.  Again, we are 
just giving some hints about methods that need to be further developed.  This is analogous to how malleable cryptographic
functions are made non-malleable by composing them with other
suitable functions.

\section{Dialects}\label{sec:dialects}
\begin{figure}
    \centering
    \includegraphics[width=\linewidth]{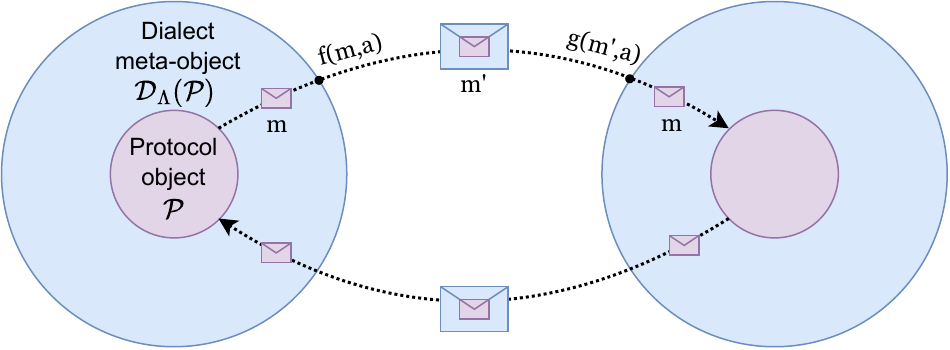}
    \caption{Dialect meta-object enhanced from \cite{ProtocolDialects-FormalPatterns}.}
    \label{fig:dialect-meta-objects}
\end{figure}

The \emph{dialect} construction is a \emph{formal pattern} of the form :
\[
\mathcal{D}: (\lingo(D), \proto(D)) \mapsto \dlct(\lingo(D), \proto(D))
\]
where $\proto(D)$ is a protocol, i.e., an actor-based system, with $D$ its data type of payloads, $\lingo(D)$ is a
lingo with input data type $D$, and  
$\dlct(\lingo(D), \proto(D))$ is a transformed protocol that endows $\proto(D)$ with lingo $\lingo(D)$.
In $\dlct(\lingo(D), \proto(D))$,
each protocol participant of $\proto(D)$ is \emph{wrapped}
inside a \emph{dialect meta-object} that uses lingo $\lingo(D)$ to obfuscate the 
communication between the honest protocol actors of $\proto(D)$.
We call $\dlct(\lingo(D), \proto(D))$ \emph{a dialect} for $\proto(D)$.
The $\mathcal{D}$ construction is both \emph{protocol-generic} and \emph{lingo-generic}: it applies to any protocol $\proto(D)$
and to any lingo $\lingo(D)$, provided $D$ is shared.
The essential behavior of a dialect is summarized
in Figure \ref{fig:dialect-meta-objects}, which shows how a meta-object's sending action uses $f$ to
transform outgoing messages and a meta-object's receiving action uses $g$ (both with the same \emph{shared} secret parameter $a$)
to decode the original payload and pass it to the receiving object.  In $\dlct(\lingo(D), \proto(D))$  the code
of protocol $\proto(D)$ is not changed in any manner: the meta-object wrappers are added in
an entirely modular way.
%

%
%

In Maude, the generic $\mathcal{D}$ construction is specified as parametrised object modules (using keywords \texttt{omod} and \texttt{endom}), with two parameter theories: \texttt{PMLINGO} and \texttt{PROTOCOL}. The former is the theory for Lingos presented in Section~\ref{sec:lingos}. 
%
%
The \texttt{PROTOCOL} theory is specified as    
\begin{Verbatim}
oth PROTOCOL is 
    sort Payload . msg to_from_:_ : Oid Oid Payload -> Msg . 
endoth
\end{Verbatim}
requiring a sort \texttt{Payload}, which has to match with sort \texttt{D1} from \texttt{PMLINGO}, and a message format. The object module also declares a $Dialect$ class with the following attributes.

\begin{enumerate}
    \item \texttt{conf}: the underlying protocol (or dialect)
    \item \texttt{in-buffer}: collection of incoming messages before applying the lingo's decoding function (i.e. $g$)
    \item \texttt{peer-counters}: a map between object identifiers and natural numbers that serves as a counter for the number of messages exchanged. It is also used in the computation of the parameters used by the lingo's transformation functions.
\end{enumerate}

Furthermore, dialects add three new rules available in Figure~\ref{fig:wrapper-actor-rules}. These rules are protocol generic, meaning they can be applied independently of the protocol. The rule labels and their semantics are as follows.

\begin{enumerate}
    \item Rule labelled $\operatorname{out}$ processes messages produced by the wrapped protocol, sending the transformed messages into the network, by applying the lingo's $f$ function with the corresponding parameter.
    \item Rule labelled $\operatorname{deliver}$ collects incoming messages to the underlying protocol in the \texttt{in.buffer} attribute.
    \item Rule labelled $\operatorname{in}$ processes messages stored in the buffer, by applying the lingo's $g$ function with the proper parameter, delivering the original version of the message to the protocol.
\end{enumerate}

Notice how in Figure~\ref{fig:wrapper-actor-rules}, both rules \texttt{out} and \texttt{in} apply the f and g respectively and supply the argument by calling the lingo's \texttt{param} function. Recall from Section~\ref{sec:lingos}, that this function computes a value $a \in A$ from $\mathbb{N}$. The dialect meta-object is responsible of making the parameter change. In a nutshell, the dialect supplies to the \texttt{param} function the current message number from the map \texttt{peer-counter}. Nevertheless, this is one of the approaches that dialects can use to make lingo's more unpredictable.

\begin{figure*}[ht!]
\begin{align*}
\mathit{in}\; &:\; < O_1 : \textit{DC} \mid \mathit{in.buffer} : M_{in}, \mathit{atts}'>\ (to \ O_1 \ from \ O_2 \ : \ P) \\
\rightarrow\ &< O_1 : \textit{DC} \mid \mathit{in.buffer} : (M_{in} \cup (to \ O_1 \ from \ O_2 \ : \ P), \mathit{atts}' > \ 
\\[1ex]
\mathit{out}\; &:\;
< O_1 : \mathit{DC} \mid \mathit{conf} : (< O_1 : C \mid \mathit{atts} >\ (to \ O_2 \ from \ O_1 \ \texttt{:} \ P) \cup \ M),\ \\&\quad\quad\quad\quad\quad\quad\;\mathit{peer.counters}:R ,\ \mathit{atts}'> \\
\rightarrow&\ < O_1 : \mathit{DC} \mid \mathit{conf} : (< O_1 : C \mid \mathit{atts} >\ M \setminus (to \ O_2 \ from \ O_1 \ : \ P)),\\
&\quad\quad\quad\quad\quad\quad\operatorname{update}(\mathit{peer.counters}:R, \mathit{atts}') > \\
&\quad (to \ O_2 \ from \ O_1\ : \ f(P, param(R[O_2]))) 
\\[1ex]
\mathit{deliver}\; &:\;
< O_1 : \textit{DC} \mid \mathit{conf} : (< O_1 : C \mid \mathit{atts} > M),\ \mathit{in.buffer} : M_{in},\\
&\quad\quad\quad\quad\quad\quad\;\mathit{peer.counters}:R ,\ \mathit{atts}'> \\
\rightarrow&\ < O_1 : \textit{DC} \mid \mathit{conf} : (< O_1 : C \mid \mathit{atts} > M \cup \{g(M_{selected}, \operatorname{param}(R[O_2]))\},\\ 
&\quad\quad\quad\quad\quad\quad\mathit{in.buffer} : (M_{in} \setminus M_{selected}),\ \\
&\quad\quad\quad\quad\quad\quad {update}(\mathit{peer.counters}:R, \mathit{atts}') > \\
\mathit{if}&\ size(M_{in}) \geq egressArity \; \land M_{selected} := take(egressArity, M_{in}) \; \land\\
&fromOidTag(M_{selected}, R) = true \; \texttt{.}
\end{align*}
\vspace{-4ex}
\caption{Meta-Actor Rewrite Rules.}
\label{fig:wrapper-actor-rules}
\vspace{-2ex}
\end{figure*}


   
\begin{example}[Dialects over MQTT]\label{ex:dialect}
    Given the lingo from Example~\ref{ex:XOR-BSeq-Lingo} we can transform the MQTT protocol of Example~\ref{ex:MQTT-formal-semantics} by means of a dialect. To do so we need to instantiate the \texttt{DIALECT} module with a valid lingo view and a protocol view such as

    \begin{Verbatim}
protecting DIALECT{xor-seq-l, MqttProtocol{Nat} .
    \end{Verbatim}

    Let us use the the \texttt{BitVec\{N\}} type described in Section~\ref{sec:lingos} as the payload of messages from MQTT. We can have obtain a dialect where the lingo uses the exclusive or operation over \texttt{BitVec\{N\}}, but adapted to comply with the \texttt{PMLINGO} theory. Our formal semantics of MQTT is parametric enough to provide a representation of its payload as bit vectors just by changing the view given to the parameter. The instantiation of such a scenario is as follows.
    
    \begin{Verbatim}
protecting DIALECT{xor-lingo{BitVec{4}}, MqttProtocol{BitVec{4}}} .
    \end{Verbatim}

    In the case we need to increase the size, for example to 8, we can do so in a very simple way. Just by changing the view given as parameter, for both lingo and protocol, we obtain a more secure dialect 
    since the range of parameters $A$ has increased exponentially.
   
    \begin{Verbatim}
protecting DIALECT{xor-lingo{BitVec{8}}, MqttProtocol{BitVec{8}}} .
    \end{Verbatim}

    Another dialect instantiation with stronger guaranties would be the lingo from Example~\ref{ex:horizontal-comp}. We provide in Appendix~\ref{app:zip} an adapted example view \texttt{xor\&dc-lingo-BitVec\{8\}} for Example~\ref{ex:horizontal-comp} and a view \texttt{dhor-comp} for \texttt{DHOR-COMP}. Thanks to the modularity of lingos we can obtain the new dialect just with the following line.

    \begin{Verbatim}
protecting DIALECT{DHOR-COMP{xor&dc-lingo-BitVec{8}}, 
                             MqttProtocol{BitVec{8}}} .
    \end{Verbatim}

    Furthermore, if we want a dialect in which messages are first encoded with XOR, and then made stronger through the use of Divide and Check, we can automatically obtain a dialect that uses the lingo of Example~\ref{ex:functional-comp} as
        
    \begin{Verbatim}
protecting DIALECT{FUN-COMP{xor+dc-lingos-BitVec{8}}, 
                            MqttProtocol{BitVec{8}}} .
    \end{Verbatim}

    All the examples shown in the paper are executable in Maude, and we made them available in a GitHub repository at \url{https://github.com/v1ct0r-byte/Protocol-Dialects-as-Formal-Patterns}. 
\end{example}

\section{Concluding Remarks}\label{sec:conclusion}

Dialects are a resource-efficient approach as a first line of defense against outside attackers. Lingos are invertible message transformations used by dialects to 
thwart attackers from maliciously disrupting communication. In this paper we propose new kinds of lingos and make them stronger through lingo transformations, 
including composition operations, all of them new. 
These lingo transformations and composition
operations are \emph{formal patterns} with desirable security properties.
We also propose a refined and simpler definition for dialects by replacing two former dialect composition operations in \cite{ProtocolDialects-FormalPatterns} by considerably 
simpler and more efficient lingo composition operations.  
All these concepts and transformations have been formally specified and made executable in Maude.

Our next step is to explore lingos in action as they would be used by dialects in the face of an on-path attacker.  We have been developing a formal intruder model that, when combined with the dialect specifications provided in this paper, can be used for statistical and probabilistic model checking of dialects. This model provides the capacity to specify the probability that the intruder correctly guesses secret parameters.
For this we are using QMaude\cite{rubio2023qmaude}, which supports probabilistic and statistical model checking analysis of Maude specifications.



\bibliographystyle{ieeetr}
\bibliography{biblio/dialects, biblio/maude}

\end{document}